\documentclass[preprint]{aastex}	

\newcommand{\hi}{\hbox{\ion{H}{1}}}	
\newcommand{\hii}{\hbox{\ion{H}{2}}}    

\newcommand{\lsun}{\hbox{L$_{{\odot},B}$}}
\newcommand{\msun}{\hbox{M$_{\odot}$}}

\newcommand{\otwo}{\hbox{[\ion{O}{2}]$\lambda$ 3727}}
\newcommand{\nethree}{\hbox{[\ion{Ne}{3}]$\lambda$ 3869}}
\newcommand{\hdelta}{\hbox{H$\delta$}}
\newcommand{\hgamma}{\hbox{H$\gamma$}}
\newcommand{\othreea}{\hbox{[\ion{O}{3}]$\lambda$ 4363}}
\newcommand{\hbeta}{\hbox{H$\beta$}}
\newcommand{\othree}{\hbox{[\ion{O}{3}]$\lambda\lambda$ 4959,5007}}

\newcommand{\othreec}{\hbox{[\ion{O}{3}]$\lambda$ 5007}}

\newcommand{\halpha}{\hbox{H$\alpha$}}
\newcommand{\ntwob}{\hbox{[\ion{N}{2}]$\lambda$ 6583}}
 
\newcommand{\stwo}{\hbox{[\ion{S}{2}]$\lambda\lambda$ 6716,6731}}

\newcommand{\ntwootwo}{\hbox{[\ion{N}{2}]/[\ion{O}{2}]}}

\shortauthors{Lee \& Skillman}
\shorttitle{H II Region Abundances in NGC 1705}
\slugcomment{\sc ApJ, accepted} 

\begin{document}

\title{
Chemical Abundances of \ion{H}{2} Regions in the
Starburst Galaxy NGC 1705$\,$\altaffilmark{\dag}
}
\author{
Henry Lee and Evan D. Skillman
}
\affil{
Department of Astronomy, University of Minnesota, \\
116 Church St. SE, Minneapolis, MN 55455 USA \\
{\tt hlee@astro.umn.edu, skillman@astro.umn.edu}
}

\altaffiltext{\dag}{
Based on EFOSC2 observations collected at the European Southern
Observatory, Chile.
}

\begin{abstract}			
We report optical spectroscopy of 16 
\hii\ regions in NGC~1705 and \othreea\ detections for the
first time in five \hii\ regions.
The resulting mean oxygen abundance derived directly from 
measured electron temperatures is 12$+$log(O/H) = 
$8.21 \pm 0.05$, which corresponds to 
[O/H] = $-$0.45, or 35\% of the solar value.
There are no significant spatial inhomogeneities in \othreea\ oxygen
abundances from \hii\ regions at a radius approximately
10\arcsec\ from the super star cluster.
In \hii\ regions where \othreea\ was not measured, oxygen abundances
derived with bright-line methods (accurate only to 0.2~dex) are in
agreement with direct values of the oxygen abundance.
Faint narrow He~II$\,\lambda$ 4686 emission is found in two 
\hii\ regions, but the implied contribution from O$^{+3}$ to the 
total oxygen abundance is only 0.01~dex.
The mean argon-, neon, and nitrogen-to-oxygen abundance ratios 
are consistent with mean values for other dwarf irregulars, blue
compact dwarf galaxies, and \hii\ galaxies at comparable
oxygen abundances.
Interestingly, the nitrogen-to-oxygen abundance ratio in the ionized
\hii\ gas agrees with the value for the neutral \hi, even though
the metallicity of the neutral gas may be a factor of six lower
than that of the ionized gas.
This may be indicative of low-metallicity gas in the halo
of the galaxy.
Extinction values, $A_V$, derived from observed Balmer line
ratios along lines of sight to \hii\ regions are in the range between
zero and 0.9~mag.
Significant and variable extinction may have important effects on the
interpretation of resolved stellar populations and derived star
formation histories.
With respect to the metallicity-luminosity and metallicity-gas
fraction diagnostics, the measured oxygen abundance for 
NGC~1705 is comparable to Local Group dwarf irregulars at
a given luminosity and gas fraction.
Simple chemical evolution models suggest that the galaxy is
quickly evolving into a gas-poor dwarf galaxy.
\end{abstract}


\section{Introduction}

Dwarf galaxies play a key role in our understanding of galaxy
formation and evolution, star formation, and cosmology.
These are thought to be the primary building blocks in hierarchical
models of structure formation (e.g., \citealp{kfg93}).
Dwarf galaxies have low masses and, typically, low metallicities
(below half-solar; e.g., \citealp{mateo98}).
However, there is a range of morphologies, and recent attention
has been paid to understanding the relationships between different
types of dwarf galaxies and/or whether or not they all have a common
progenitor.
Dwarf irregular galaxies could be spending most of their lives in a
relatively quiescent phase by forming stars at a low constant rate
(e.g., IC~1613, \citealp{evan_ic1613}).
However, blue compact dwarf galaxies appear to host recent bursts of
star formation, whose optical luminosities are often larger than the
underlying stellar populations.
Blue compact dwarfs then may be related to gas-poor dwarf spheroidal
galaxies, if the latter experienced in the past strong burst(s) of
star formation, which eventually blew out their gas content.
Because of their low gravitational potentials, dwarf galaxies are
expected to lose gas and metals in supernova-driven winds or outflows.
These winds may be an important mechanism to the chemical enrichment
of the intergalactic medium (e.g., \citealp{garnett02}).

Starbursts offer a unique glimpse into a distinct phase in the life of
a dwarf galaxy, clues to the possible evolutionary relationship
between different types of dwarf galaxies, and insights into the
processes regarding the formation of massive stars.
In galaxies with strong starbursts, star formation occurs at a rate
much larger than the past average rate.
The physical conditions within starburst regions are thought to be
similar to conditions in the early universe, where the first
generations of stars formed and galaxy assembly occurred.
Recent ultraviolet studies, e.g., with the Far Ultraviolet
Spectroscopic Explorer ({\em FUSE}), have recently
revolutionized the field, because the ultraviolet probes directly the
hot and massive star content in the starbursts.
Optical emission-line spectra of \hii\ regions are dominated by strong
narrow emission lines, a relatively weak continuum which rises
towards blue wavelengths, and weak absorption Balmer-line features. 
These spectra provide diagnostics on the conditions within the ionized
nebular gas, information about the total reddening along various lines of
sight, and information about the most recent chemical enrichment in
the form of elemental abundances and ratios (e.g., \citealp{kd02}).  
In fact, the metallicity may be the most important parameter in
determining the properties of starburst galaxies at ultraviolet
wavelengths (e.g., \citealp{heckman98,tremonti01}). 

\subsection{NGC 1705}

NGC~1705 is a nearby, isolated, amorphous or blue compact
dwarf galaxy, which has been the subject of much recent attention in
the literature.
A basic list of properties is given in Table~\ref{table_n1705}.
Near the center is a $10^5$ \msun\ super star cluster (SSC),
also known as NGC~1705-1 \citep{mmt85,meurer89,meurer92}, whose
properties appear to be similar to Galactic globular clusters
\citep{hf96}. 
However, the SSC is not situated exactly at the center of the galaxy
\citep{tosi01}, which was previously suggested by \cite{hl97} and
\cite{hensler98}. 
\cite{tosi01} showed that stars more massive than about 30~\msun\ and
younger than 20~Myr were located within 4\arcsec\ of the SSC.
The most recent episode of star formation appears to have been 
dominated by the SSC, whose age has been measured to be in the
range 10 to 15 Myr \citep{mmt85,oconnell94,hl97,vazquez04}.
The SSC is surrounded by clusters of young massive stars, which
are seen as bright \hii\ regions 
(Fig.~\ref{fig_n1705ha}; e.g., 
\citealp{mmt85,meurer89,meurer92,hhg93,marlowe95,gildepaz03}).
The underlying stellar population has been shown to be a composite mix
of stars ranging in age from 0.1 to 10~Gyr
\citep{meurer92,qrf95,annibali03}. 
\cite{greve96} detected no CO emission at the position of the SSC.
Using {\em FUSE\/} to look for far-ultraviolet absorption lines
of $H_2$, \cite{hoopes04} measured the central 30\arcsec\ by 30\arcsec\
in NGC~1705 and set an upper limit of $3.90 \times 10^{14}$~cm$^{-2}$
for the total $H_2$ column density.

NGC~1705 is thought to be a prototype for dwarf galaxies presently
undergoing mass loss or outflows.
Evidence from previous work is summarized here in chronological
order.
\cite{lamb85} observed the galaxy with the International Ultraviolet
Explorer ({\em IUE\/}), and noted that the galaxy was going through 
an overall modest enhancement of star formation, which was superposed
on the center undergoing a post-starburst phase.
\cite{york90} also used {\em IUE\/} data and proposed that absorption
lines seen in their ultraviolet spectra (e.g., \ion{Al}{2}
$\lambda$ 1671~\AA) arose from intervening absorption systems in the
interstellar medium of the galaxy.
\cite{meurer92} published a comprehensive study combining imaging and
spectroscopy from ultraviolet to radio wavelengths. 
A complex set of arc and loop features seen in \halpha\ roughly
centered on the SSC suggested that explosive events had clearly
occurred in the system, and that the SSC was responsible for creating
a very large, bright, central ``superbubble'' of ionized
gas.\footnote{
With {\em HST\/} narrow-band imaging of four starburst galaxies, 
\cite{calzetti04} showed that the central starbursts produce
sufficient mechanical energy from shocks arising from massive star
winds and supernova explosions to explain the nonphotoionized gas
content and that the nonphotoionized gas is associated with extended
shells or filaments.
} 
They also showed that line profiles in \halpha\ exhibited multiple
components with radial velocity differences of about 100 km~s$^{-1}$
across the center of the galaxy. 
The largest blueshifted component corresponded to a strong
interstellar absorption feature, which suggested that this component
or feature was expanding and situated in front of the SSC. 
\cite{hhg93} also confirmed the existence of large \halpha\ filaments
centered on the SSC.
With Fabry-Perot and echelle spectroscopic data, \cite{marlowe95}
showed from \halpha\ kinematics and profiles that the superbubble was
expanding and that the dynamical timescale was of order 10~Myr.
Among amorphous galaxies, \cite{marlowe97} classified NGC~1705 as
a galaxy with ``a strong core and explosive \halpha\ morphology.''
With {\em ROSAT}, \cite{hensler98} found embedded among the \halpha\
arcs two regions of soft X-ray emission.
These structures suggested that the X-ray pockets were surrounded
by cool gas shells in \halpha\ emission, and that these features could
be interpreted as being the outer plumes of the same superbubble. 
Using Goddard High-Resolution Spectrograph (GHRS) archival data from
the Hubble Space Telescope ({\em HST\/}), \cite{sb97} and
\cite{sahu98} reported three absorption-line systems from ultraviolet
absorption lines.
One of the systems with the largest line-of-sight velocity was
associated with a blueshifted emission component of the
expanding supershell.
The absorption system contained strong \ion{Si}{2} and 
\ion{Al}{2} absorption, but weak \ion{Fe}{2} $\lambda$ 1608
absorption.
\cite{heckman01} measured \ion{O}{6} outflow with {\em FUSE}, and
suggested that the warm and hot gas phases were flowing out of the
galaxy.
\cite{johnson03} did not detect any thermal radio sources within
NGC~1705, which supported the idea that the galaxy is currently
in a post-starburst phase.
This would mean that at the present time, any massive star formation
has somehow been completely shut off in the galaxy, even though a
strong starburst event occurred $\approx$ 10~Myr ago. 

The SSC has likely been the source responsible for expelling the
outflowing gas in the last $\approx$ 10~Myr (e.g., 
\citealp{oconnell94,suchkov94,hl97,tosi01}).
Naturally, the SSC was also thought to be the main source of
ionization in the galaxy. 
However, \cite{vazquez04} showed that the SSC could not be the
dominant ionizing source in the galaxy; instead, O-type stars within 
\hii\ regions surrounding the SSC are more likely to be the sources 
of the ionizing radiation.

For the metallicities typical of dwarf galaxies, oxygen is 
a relatively abundant element, and regulates physical conditions
as the primary coolant within \hii\ regions
(e.g., \citealp{dinerstein90,skillman98}).
The cooling is accomplished primarily through the optical emission
lines of oxygen. 
Oxygen abundances from \hii\ regions can help constrain the modeling
required to extract the metallicity in the hot, ejected, X-ray
emitting gas, which is expected to appear in galaxy halos (e.g.,
\citealp{martin02}). 
The nebular oxygen abundance for NGC~1705 was thought to be comparable
to that of the LMC (12$+$log(O/H) = 8.35; \citealp{rd92}).
However, previous determinations of oxygen abundances have been
derived in the absence of measurements for the temperature-sensitive
\othreea\ emission line \citep{lamb85,meurer92,sb94,heckman98}.
Oxygen abundances derived using bright-line or empirical methods are
accurate to within 0.2~dex (\S~\ref{sec_oxybrightlines}; also 
\citealp{lee03south}; \citealp{scm03b}).
As part of an ongoing program to understand the nature of nearby
starburst galaxies, we acquired spectra of \hii\ regions in NGC~1705 
in order to detect the \othreea\ line, to derive subsequently electron
temperatures and direct oxygen abundances, and to measure extinction
values using Balmer line ratios along different lines of sight.

The paper is organized as follows.
Observations and reductions of the data are presented in
\S~\ref{sec_obs}, 
the measurements are given in \S~\ref{sec_measurements},
derivations of chemical abundances are described in
\S~\ref{sec_abund}, 
discussion of the results is presented in \S~\ref{sec_discuss},
and the conclusions are given in \S~\ref{sec_concl}.
For the remainder of this paper, we adopt 12$+$log(O/H) = 8.66 as the
revised solar value for the oxygen abundance, and $Z_{\odot}$ = 0.0126
as the revised solar mass fraction in the form of metals
\citep{asplund04}.

\section{Observations and Reductions}
\label{sec_obs}		

Long-slit spectroscopic observations were carried out during new moon 
on 2003 Aug. 27, 28, and 30 UT with the 
ESO Faint Object Spectrograph and Camera (EFOSC2) 
instrument on the 3.6-m telescope at ESO La Silla Observatory. 
Conditions were mostly cloudy on Aug. 30, and the data obtained 
were not used. 
Details of the instrumentation employed and the log of observations
are listed in Tables~\ref{table_obsprops} and \ref{table_obslog},
respectively.
The \halpha\ images published by \cite{gildepaz03} were used to
optimize slit placements on the galaxy.
Two-minute \halpha\ acquisition images were obtained in order to set
the best position angle of the slit, so that the slit could go through
as many \hii\ regions possible.
Data were obtained with three slit placements; the position angles are
listed in Table~\ref{table_obslog}.
Actual slit orientations on the galaxy are shown in
Figs.~\ref{fig_n1705ha} to \ref{fig_slitc} inclusive.
\hii\ regions for which spectra were acquired are given in
Table~\ref{table_h2reglist}.
The locations of \hii\ regions were compared and matched with
identifications in the images published by \cite{mmt85} and
\cite{meurer92}.

The spectra were reduced using standard IRAF\footnote{
IRAF is distributed by the National Optical 
Astronomical Observatories, which is operated by the Associated
Universities for Research in Astronomy, Inc., under contract to the
National Science Foundation.}
routines.
Data obtained on a given night were reduced independently.
The raw two-dimensional images were subtracted for bias and trimmed.
Dome flat exposures were used to remove pixel-to-pixel variations 
in response, and 
twilight flats were acquired at dusk each night to correct
for variations over larger spatial scales.
To correct for the ``slit function'' in the spatial direction, the
variation of illumination along the slit was taken into account
using dome and twilight flats. 
Cosmic rays were removed in the addition of multiple exposures.
For the single frames taken with slits B and C, the spectra were
inspected visually for cosmic rays.
Wavelength calibration was obtained using helium-argon (He-Ar) arc
lamp exposures taken throughout each night.
With the 5\arcsec\ slit, exposures of standard stars Feige~110,
G138$-$31, LTT~1788, LTT~7379, and LTT~9491 were used for flux
calibration.
The flux accuracy is listed in Table~\ref{table_obslog}. 
One-dimensional spectra for each \hii\ region were obtained 
with unweighted summed extractions.  
Spectra for which \othreea\ was detected are shown in
Fig.~\ref{fig_specall}.
For completeness, the spectrum of the SSC with slit C is shown in
Fig.~\ref{fig_ssc}.
We observed weak \othreec\ and \halpha, as seen previously by
\cite{mmt85}.
The spectrum of the SSC is not discussed further in the remainder of 
this paper.

\section{Measurements}
\label{sec_measurements}

Emission-line strengths were measured using software developed
by M. L. McCall and L. Mundy; see \cite{lee01,lee03field,lee03virgo}.
Because the systemic velocity for NGC~1705 is over 600 km~s$^{-1}$,
there is sufficient wavelength offset from the Hg~I $\lambda$ 4358 sky
line for the \othreea\ emission line to be detected.
The spectra clearly indicate the presence of \othreea\ in five 
\hii\ regions: A3, B3, B4, B6, and C6.

Corrections for reddening and for underlying absorption 
and abundance analyses were performed with SNAP 
(Spreadsheet Nebular Analysis Package, \citealp{snap97}). 
Balmer fluxes were first corrected for underlying Balmer absorption
with an equivalent width 2~\AA\ \citep{mrs85,lee03virgo}.
Here, the uncertainty in the correction for underlying Balmer
absorption was assumed to be zero (but see also the discussion
below).
\halpha\ and \hbeta\ fluxes were used to derive
reddening values, $E(B-V$), using the equation
\begin{equation}
\log\frac{I(\lambda)}{I(\hbeta)} = 
\log\frac{F(\lambda)}{F(\hbeta)} + 
0.4\,E(B-V)\,\left[A_1(\lambda) - A_1(\hbeta)\right]
\label{eqn_corrthesis}
\end{equation}
\citep{lee03field}.
$F$ and $I$ are the observed flux and corrected intensity ratios,
respectively.
Intrinsic case-B Balmer line ratios determined by \cite{sh95} were
assumed.
$A_1(\lambda)$ is the extinction in magnitudes for $E(B-V) = 1$, 
i.e., $A_1(\lambda) = A(\lambda)/E(B-V)$, where 
$A(\lambda)$ is the monochromatic extinction in magnitudes.
Values of $A_1$ were obtained from the \cite{cardelli89}
reddening law as defined by a ratio of the total to selective
extinction, $R_V$ = $A_V/E(B-V)$ = 3.07, which in the limit of zero
reddening is the value for an A0V star (e.g., Vega) with intrinsic
color $(B-V)^0 = 0$.
Because \stwo\ lines were generally unresolved, 
$n_e$ = 100~cm$^{-3}$ was adopted for the electron density.
Errors in the derived $E(B-V)$ were computed from the maximum and
minimum values of the reddening based upon $2\sigma$ errors in the
fits to emission lines. 

Observed flux $(F)$ and corrected intensity $(I)$ ratios are listed
in Tables~\ref{table_data1} to \ref{table_data3} inclusive.
The listed errors for the observed flux ratios at each wavelength
$\lambda$ account for the errors in the fits to the line profiles,
their surrounding continua, and the relative error in the sensitivity
function stated in Table~\ref{table_obslog}.  
Errors in the corrected intensity ratios account for maximum and
minimum errors in the flux of the specified line and of 
the \hbeta\ reference line.
At the \hbeta\ reference line, errors for both observed and corrected
ratios do not include the error in the flux.
Also given for each \hii\ region are: the observed \hbeta\ flux, the
equivalent width of the \hbeta\ line in emission, and the derived 
reddening from SNAP.

Where \othreea\ is measured, we also have performed the additional 
computations to check the consistency of our results.
Equation~(\ref{eqn_corrthesis}) can be generalized and rewritten as
\begin{equation}
\log\frac{I(\lambda)}{I(\hbeta)} =
\log\frac{F(\lambda)}{F(\hbeta)} + c(\hbeta) \, f(\lambda),
\label{eqn_corr}
\end{equation}
where $c(\hbeta)$ is the logarithmic extinction at \hbeta, and 
$f(\lambda)$ is the wavelength-dependent reddening function 
\citep{aller84,osterbrock}.
From Equations~(\ref{eqn_corrthesis}) and (\ref{eqn_corr}), 
we obtain
\begin{equation}
c(\hbeta) = 1.43 \, E(B-V) = 0.47 \, A_V.
\label{eqn_chbebv}
\end{equation}
The reddening function normalized to \hbeta\ is derived from
the \cite{cardelli89} reddening law, assuming $R_V$ = 3.07.
As described in \cite{scm03b}, values of $c(\hbeta)$
were derived from the error weighted average of values for
$F(\halpha)/F(\hbeta)$, $F(\hgamma)/F(\hbeta)$, and
$F(\hdelta)/F(\hbeta)$ ratios while simultaneously solving for the
effects of underlying Balmer absorption with equivalent
width, EW$_{\rm abs}$.
We assumed that EW$_{\rm abs}$ was the same for \halpha, \hbeta,
\hgamma, and \hdelta.
Uncertainties in $c(\hbeta)$ and EW$_{\rm abs}$ were determined
from Monte Carlo simulations \citep{os01,scm03b}.
Errors derived from these simulations are larger than errors
quoted in the literature by either assuming a constant value for
the underlying absorption or derived from a $\chi^2$ analysis in
the absence of Monte Carlo simulations for the errors; 
Fig.~\ref{fig_monte} shows an example of these simulations for \hii\
region A3.
In Tables~\ref{table_data1} to \ref{table_data3}, we included
the logarithmic reddening and the equivalent width of the underlying
Balmer absorption, which were solved simultaneously.
Values for the logarithmic reddening are consistent with values
of the reddening determined with SNAP.
Where negative values were derived, the reddening was set to zero
in correcting line ratios and in abundance calculations.

Relatively narrow \ion{He}{2} emission was detected in \hii\ regions 
B3 and B4.
In \hii\ region B3, the equivalent width of the \ion{He}{2} $\lambda$
4686 emission line is $1.71 \pm 0.20$~\AA\ (Fig.~\ref{fig_he2};
Table~\ref{table_data2}); the line is somewhat weaker in B4
(equivalent width $0.77 \pm 0.11$~\AA).
Using models for young stellar populations in starbursts by
\cite{sv98}, our \ion{He}{2} data are best fit by a model
with $Z = 0.008$, a Salpeter stellar initial mass function with upper
mass limit of 120~\msun, and an instantaneous burst of star
formation at an age of about 5~Myr.
One can compare our data with the broad \ion{He}{2} emission seen, for
example, in the Sculptor group starburst galaxy NGC~625 
\citep{scm03b}.

\section{Nebular Abundances}
\label{sec_abund}

Oxygen abundances in \hii\ regions were derived using three methods:
(1) the direct method (e.g., \citealp{dinerstein90,skillman98});
and the bright-line methods discussed by
(2) \cite{mcgaugh91}, which is based on photoionization models;  
and (3) \cite{pilyugin00}, which is purely empirical.

\subsection{Oxygen Abundances: \othreea\ Temperatures}

For the ``direct'' conversion of emission line intensities into ionic
abundances, a reliable estimate of the electron temperature in the
ionized gas is required.
\hii\ regions are modeled as two zones: a low- and a high-ionization
zone, characterized by temperatures $T_e($O$^+)$ and $T_e($O$^{+2})$,
respectively. 
The temperature in the O$^{+2}$ zone is measured with
the emission line ratio $I$(\othreec)/$I$(\othreea) 
\citep{osterbrock}.
The temperature in the O$^+$ zone is given by 
\begin{equation}
t_e([{\rm O\;II}]) = 0.7 \, t_e([{\rm O\;III}]) + 0.3,
\label{eqn_toplus}
\end{equation}
where $t_e = T_e/10^4$~K \citep{ctm86,garnett92}.
Within SNAP, the uncertainty in $T_e($O$^{+2})$ is computed from the
maximum and minimum values derived from the uncertainties in corrected 
emission line ratios.
The computation does not include uncertainties in the reddening (if any),
the uncertainties in the atomic data, or the presence of temperature
fluctuations.
The uncertainty in $T_e$(O$^+$) is assumed to be the same
as the uncertainty in $T_e$(O$^{+2}$).
These temperature uncertainties are conservative estimates, and are
likely overestimates of the actual uncertainties.
For subsequent calculations of ionic abundances, we assume
the following electron temperatures \citep{garnett92}:
$t_e$(N$^+$) = $t_e$(O$^+$), 
$t_e$(Ne$^{+2}$) = $t_e$(O$^{+2}$), and
$t_e$(Ar$^{+2}$) = 0.83 $t_e$(O$^{+2}$) $+$ 0.17. 

The total oxygen abundance by number is given by
O/H = O$^0$/H $+$ O$^+$/H $+$ O$^{+2}$/H $+$ O$^{+3}$/H.
For conditions found in typical \hii\ regions and those presented
here, very little oxygen in the form of O$^0$ is expected, and is not
included here.
Ionic abundances for O$^+$/H and O$^{+2}$/H were computed using
O$^+$ and O$^{+2}$ temperatures, respectively, as described above.
\ion{He}{2}$\lambda$ 4686~\AA\ emission is indicative of the presence of
O$^{+3}$, which is generally a small contributor to the total
oxygen abundance.
For example, \cite{ks93} measured He~II emission in I~Zw~18 and found
that the resulting O$^{+3}$ contribution was of order one to four percent.
In \hii\ region B3, \ion{He}{2}$\lambda$ 4686~\AA\ emission is 
about five percent of \hbeta. 
The resulting contribution by O$^{+3}$ to the total oxygen abundance
is of order 5\% or 0.01~dex.
Thus, the O$^{+3}$ contribution is small and is not included in the
total oxygen abundance.

Measurements of the \othreea\ line were obtained and subsequent 
electron temperatures were derived in five of the 16 \hii\ regions.
Ionic abundances and total abundances are computed using the 
method described by \cite{lee03field}. 
With SNAP, oxygen abundances were derived using the five-level atom
approximation \citep{fivel}, and transition probabilities and collision
strengths for oxygen from \cite{pradhan76}, \cite{mb93}, and 
\cite{wiese96}. 
Balmer line emissivities from \cite{sh95} were used. 
Derived ionic and total abundances are listed in
Tables~\ref{table_abund1} and \ref{table_abund2}.
These tables include derived O$^+$ and O$^{+2}$ electron temperatures; 
O$^+$ and O$^{+2}$ ionic abundances and the total oxygen abundances.
Errors in direct oxygen abundances computed with SNAP have two
contributions: the maximum and minimum values for abundances from
errors in the temperature, and the maximum and minimum possible values
for the abundances from propagated errors in the intensity ratios. 
These uncertainties in oxygen abundances are also conservative
estimates. 

Using the method described by \cite{scm03b}, we recompute oxygen
abundances in \hii\ regions with \othreea\ detections.
Abundances are computed using the emissivities from the five-level
atom program by \cite{sd95}.
As described above, we use the same two-temperature zone model and 
temperatures for the remaining ions.
The error in $T_e$(O$^{+2}$) is derived from the uncertainties in
the corrected emission-line ratios, and does not include any
uncertainties in the atomic data, or the possibility of temperature
variations within the O$^{+2}$ zone.
The fractional error in $T_e$(O$^{+2}$) is applied similarly to
$T_e$(O$^+$) to compute the uncertainty in the latter.
Uncertainties in the resulting ionic abundances are combined in
quadrature for the final uncertainty in the total linear (summed)
abundance.
The adopted \othreea\ abundances and their uncertainties computed in
this manner are listed in Tables~\ref{table_abund1} and
\ref{table_abund2}.
Direct oxygen abundances computed with SNAP are in excellent agreement
with direct oxygen abundances computed with the method described
by \cite{scm03b}; abundances from the two methods agree to within 
0.02~dex.
From the five direct oxygen abundances listed, the weighted mean is
(O/H) = $(1.62 \pm 0.19) \times 10^{-4}$, 
or 12$+$log(O/H) = $8.21 \pm 0.05$.
The mean value corresponds to [O/H] = $-$0.45~dex,\footnote{
The following standard notation is used: 
[X/H] = log(X/H) $-$ log(X/H)$_{\odot}$.
}
or 35\% of the solar value.


\subsection{Oxygen Abundances: Bright-Line Methods}
\label{sec_oxybrightlines}

For \hii\ regions without \othreea\ measurements, secondary methods
are necessary to derive oxygen abundances.
The bright-line method is so called because the oxygen abundance is
given in terms of the bright [O~II] and [O~III] lines. 
\cite{pagel79} suggested using
\begin{equation}
R_{23} = \frac{I({\otwo}) + I({\othree})}{I({\hbeta})}
\label{eqn_r23_def}
\end{equation}
as an abundance indicator.
Using photoionization models, \cite{skillman89} showed that bright
[O~II] and [O~III] line intensities can be combined to determine
uniquely the ionization parameter and an ``empirical'' oxygen
abundance in low-metallicity \hii\ regions.
\cite{mcgaugh91} developed a grid of photoionization models and
suggested using $R_{23}$ and $O_{32}$ = $I$(\othree)/$I$(\otwo) 
to estimate the oxygen abundance.
However, the calibration is degenerate such that for a given value of
$R_{23}$, two values of the oxygen abundance are possible.
The \ntwootwo\ ratio was suggested \citep{mrs85,mcgaugh94,vanzee98}
as the discriminant to choose between the ``upper branch'' (high
oxygen abundance) or the ``lower branch'' (low oxygen abundance).
In the present set of spectra, \ntwob\ line strengths are generally
small, and \ntwootwo\ values have been found to be less than the
threshold value of 0.1.
Consequently, the lower branch is used.\footnote{
Analytical expressions for the McGaugh calibration 
are found in \cite{chip99}.
}
\cite{pilyugin00} proposed a new calibration of the oxygen
abundances using bright oxygen lines.
At low abundances, his calibration is expressed as
\begin{equation}
12+\log({\rm O/H}) = 6.35 + 3.19 \log R_{23} - 1.74 \log R_3,
\label{eqn_pily}
\end{equation}
where $R_{23}$ is given by Equation~(\ref{eqn_r23_def})
and $R_3$ = $I$(\othree)/$I$(\hbeta).
In some instances, oxygen abundances with the McGaugh method could
not be computed, because the $R_{23}$ values were outside of the
effective range for the models.
\cite{scm03b} have shown that the Pilyugin calibration covers
the highest values of $R_{23}$.
Oxygen abundances derived using the McGaugh and Pilyugin bright-line
calibrations are listed in Tables~\ref{table_abund1} and
\ref{table_abund2}.
For each \hii\ region, differences between direct and bright-line
abundances are shown as a function of $O_{32}$ and $R_{23}$
in Fig.~\ref{fig_oxydiff}.
The separations between the three methods appear to increase with 
increasing $O_{32}$.
The difference between the McGaugh and Pilyugin calibrations 
(indicated by asterisks) appears to correlate with log $O_{32}$, 
which has been previously noted by \cite{scm03b} and
\cite{lee03south}. 
However, the values of $R_{23}$ in the present data are near or
at the maximum allowed in the McGaugh calibration, where the ``kink''
or the bend in the log(O/H) versus $R_{23}$ relation occurs, and
ambiguity is greatest for an estimate of the oxygen abundance
(in the absence of the \ntwootwo\ discriminant).
The bottom panel of Fig.~\ref{fig_oxydiff} shows that at these
values of $R_{23}$, the McGaugh calibration appears to give 
$\approx +0.2$ dex larger abundances than the Pilyugin calibration.
However, we remind the reader with comments from \cite{scm03b} that
the McGaugh calibration does not provide sufficient range in $R_{23}$
with his models, and the Pilyugin calibration does not include a
sufficient number of low ionization \hii\ regions.
Where \othreea\ is not measured or below the detection limit, oxygen
abundances derived using bright-line methods are in agreement with
direct abundances to within $\approx$ 0.2~dex. 
This has obvious implications to determining nebular oxygen abundances
in more distant starburst galaxies if \othreea\ cannot be detected;
see also the discussion by \cite{chip99}.

\subsection{Element Ratios}

For completeness, we derive nitrogen-to-oxygen, neon-to-oxygen,
and argon-to-oxygen ratios, which are listed in
Tables~\ref{table_abund1} and \ref{table_abund2}.
Transition probabilities and collision strengths for N$^+$, Ne$^{+2}$,
and Ar$^{+2}$ were taken from \cite{mendoza83}, \cite{mz83},
\cite{bz94}, \cite{lb94}, \cite{galavis95}, and \cite{wiese96}.

For metal-poor galaxies, it is assumed that 
N/O $\approx$ N$^+$/O$^+$ \citep{garnett90} and N$^+$/O$^+$ 
values were derived.
Nitrogen abundances were computed as N/H = ICF(N) $\times$ (N$^+$/H).
The ionization correction factor, ICF(N) = O/O$^+$, accounts for
missing ions. 
From \hii\ regions A3, B3, B4, and B6, the mean (N/O) is
$(1.76 \pm 0.21) \times 10^{-2}$, 
and the resulting log(N/O) is $-1.75 \pm 0.06$.
The poor \ntwob\ detection in \hii\ region C6 was not included in
the average.
However, if the anomalously low nitrogen value in \hii\ region B4 is
ignored, the resulting means are (N/O) = 
$(2.34 \pm 0.33) \times 10^{-2}$, and log(N/O) = $-1.63 \pm 0.07$.
Our N/O value is not significantly different from those
obtained for other dwarf galaxies at comparable oxygen abundances
\citep{ks96,vanzee97,it99}.

Neon abundances are derived as Ne/H = ICF(Ne) $\times$ (Ne$^{+2}$/H).
The ionization correction factor for neon is ICF(Ne) = O/O$^{+2}$.
For neon-to-oxygen ratios, we assumed Ne/O $\approx$
Ne$^{+2}$/O$^{+2}$ and derived a mean Ne/O of $0.374 \pm 0.028$.
We also obtain log(Ne/O) = $-0.426 \pm 0.033$, which is about 
0.3~dex higher than the average for blue compact dwarf galaxies 
\citep[$-0.72$; ][]{it99},
but our value is at the upper end of the range for \hii\ galaxies
\citep{terlevich91}.
These values could be too high, which might arise from problems with 
the reddening correction, where the $F$(\nethree)/$F$(\hbeta) 
ratio may be overcorrected. 
However, the unblended corrected Balmer line closest to \nethree\ is 
\hdelta; the H8 Balmer line is blended with an adjacent helium line,
and H$\epsilon$ is blended with [\ion{Ne}{3}] and helium lines.
Nevertheless, we find that the corrected $I$(\hdelta)/$I$(\hbeta) and
$I$(\hgamma)/$I$(\hbeta) ratios are consistent with the expected values.

Argon is more complex, because the dominant ion is not found in
just one zone.
Ar$^{+2}$ is likely to be found in an intermediate area between the
O$^+$ and O$^{+2}$ zones.
Following the prescription by \cite{itl94}, the argon abundance,
was derived as Ar/H = ICF(Ar) $\times$ Ar$^{+2}$/H.
The ionization correction factor is given by 
ICF(Ar) = Ar/Ar$^{+2}$ = $[0.15 + x(2.39 - 2.64x)]^{-1}$,
where $x$ = O$^+$/O.
From the data for \hii\ regions A3, B3, B4, B6 and C6, the mean
(Ar/O) is $(4.8 \pm 1.2) \times 10^{-3}$, and the resulting
log(Ar/O) is $-2.31 \pm 0.11$.
We also computed Ar$^{+2}$/O$^{+2}$, which agreed with the full
derivation for the argon abundance, 
i.e., Ar/O $\approx$ Ar$^{+2}$/O$^{+2}$.
Our mean log(Ar/O) is consistent with the mean determined by
\cite{it99} for their sample of blue compact dwarf galaxies.

\section{Discussion}			
\label{sec_discuss}

We present the following three points: 
(1) reddening measurements from the Balmer decrement are compared with 
previous estimates; 
(2) direct oxygen abundances derived from the present data are compared 
with previous spectroscopic measurements in the literature;
and
(3) simple chemical evolution models are used to compare the
current state of NGC~1705 with other dwarf irregulars at comparable
gas-phase metallicities.

\subsection{Reddening Estimates}

From their measure of the ultraviolet flux at 1400~\AA,
\cite{lamb85} claimed an upper limit for the reddening
internal to the galaxy, $E(B-V)_i$ $\la$ 0.2~mag.
This was consistent with reddening values derived from Balmer
decrements measured from the optical spectra of \hii\ regions in
similar amorphous galaxies, which were supposed to be relatively
dust-free.
\cite{meurer92} derived an upper limit to the internal reddening of
$E(B-V)$ $\la$ 0.02~mag from {\em IRAS\/} data.
They also detected individual features in \halpha\ profiles. 
In particular, they were able to measure blueshifted \halpha\ lines,
where the most blueshifted line was found to be at the center and
in front of the galaxy.   
Thus, they concluded that one would expect zero reddening along a line
of sight to the center of the galaxy.
\cite{calzetti94} showed that the spectral slope in the ultraviolet
continuum correlated with nebular extinction as measured from optical
Balmer emission lines.
They also found that the foreground reddening of 0.04~mag, and 
concluded that the intrinsic reddening within NGC~1705 was near zero. 
For their sample of starburst galaxies, \cite{meurer95} showed good
correlation between the far-infrared excess and the ultraviolet 
spectral slope.
This was best explained by a model geometry where most of the dust was
in a foreground screen in near proximity of the starburst, instead of
a geometry where dust was mixed with the stars; a similar conclusion
was reached by \cite{heckman98}.
Observations have indicated that the reddening towards the SSC 
was expected to be very small, which suggests that the previous
burst of star formation associated with the SSC has removed most
of the dust along our line of sight to the star cluster.
\cite{schlegel98} measured the Galactic foreground in the direction of
NGC~1705 to be reddening $E(B-V)_G$ = 0.035~mag, or an extinction in
$V$ of about 0.1~mag. 



As shown in Tables~\ref{table_data1} to \ref{table_data3}, the derived
reddenings for most of the extracted spectra are consistent with zero
or very low levels of extinction.  
In fact, all of the \hii\ regions observed in spectrum B are
consistent (within errors) with zero reddening and spectrum B covers
nebular emission over a range in galactocentric distance.  
The low levels of extinction are in agreement with previous estimates
\citep{meurer92,heckman01} which are primarily concerned with the
line of sight to the main SSC. 
However, along some lines of sight, we do detect significant
extinction. 
For example, the extinction derived for A2 is $A_V =$ 0.9~mag., and
the other \hii\ regions in that vicinity also show non-zero
reddening. 
The presence of significant and variable extinction in NGC~1705 could
be important to understanding the true nature of the history of its 
starburst.  
To date, the studies of the star formation history of NGC~1705 have
assumed a uniformly low value of extinction 
($A_V = 0.1$; \citealp{tosi01,annibali03}).  
Since the presence of variable extinction leads to a systematic bias
in the photometry of individual stars (cf.\ \citealp{cannon03}) and
variable extinction is common in other dwarf starburst galaxies 
(e.g., \citealp{calzetti99,calzetti00}), it may be interesting to
revisit the star formation history of NGC~1705 to investigate the
effect that variable extinction can have on its derived star formation
history.

\subsection{Oxygen Abundances}

\subsubsection{Comparison with Previous Results}

We compare our result with previous measurements of 
oxygen abundances published in the literature.
In their optical long-slit spectrum, \cite{lamb85} reported bright
nebular emission lines, nebular emission more extended than
the stellar core, and the center of the emission displaced with
respect to the stellar core.  
Although an oxygen abundance was not estimated, their spectrum was
suggestive of a metallicity similar to that of the LMC
(i.e., 12$+$log(O/H) = 8.35; \citealp{rd92}).
\cite{meurer92} reported for NGC~1705 an oxygen abundance of 
12$+$log(O/H) $\sim 8.46$, derived from bright emission line ratios in
their spectrum.\footnote{
Nebular diagnostics from the MAPPINGS code \citep{binette85}
were used for the derivation (G. Meurer 2004, private communication).
}
\cite{sb94} and \cite{sb95} reported gas-phase metallicities and
spectra, respectively, for 44 star-forming galaxies.
Although the placement of the long-slit on the galaxy is not known,
their spectrum of NGC~1705 showed \otwo\ and \othreec\ emission,
strong stellar continuum in the blue, but no \othreea.  
With a two-zone model, they derived $T_e$(O$^{+2}$) and $T_e$(O$^+$) 
using empirical relationships between oxygen abundance and temperature
from \cite{pagel79} and \cite{ctm86}, respectively.
The oxygen abundance was found to be 12$+$log(O/H) = 8.36.
However, with the same spectrum from \cite{sb95}, \cite{heckman98}
used the \cite{ep84} bright-line calibration, and rederived a smaller
value for the oxygen abundance: 12$+$log(O/H) = 8.0.
We have also taken the intensity ratios from \cite{sb95}, and 
we obtained 12$+$log(O/H) = 8.0 and 7.8 using the \cite{mcgaugh91}
and \cite{pilyugin00} calibrations, respectively.

The present mean value of 12$+$log(O/H) = 8.21 is approximately in the
middle of the range of reported values (8.0 to 8.5) from the literature.
When \othreea\ is too faint and/or not measured, one must rely on
bright-line methods, where the ambiguity in choosing the appropriate
``branch'' for the oxygen abundance is greatest in the range of
$R_{23}$ values considered here.
We have seen that there is a relatively broad range in oxygen
abundances spanning about 0.7~dex when various bright-line
calibrations (e.g., \citealp{pagel79,ep84,mcgaugh91,pilyugin00}) 
are considered.
The present set of \othreea\ measurements have pinned down the nebular
oxygen abundance to an accuracy of 0.1~dex.
Where \othreea\ is unmeasurable or below the detection limit, the
resulting oxygen abundances are accurate to about 0.2~dex, as
Fig.~\ref{fig_oxydiff} has shown.
Fortunately, the present results do not greatly affect the 
conclusions about the ultraviolet properties of starburst galaxies 
reached by \cite{heckman98}, or the tip of the red giant branch
distance determined by \cite{tosi01}.
\cite{tosi01} superposed stellar evolutionary tracks on their
color-magnitude diagrams, and the best agreement was reached at a
metallicity $Z = 0.004$, or about one-third solar; this is in
agreement with our value of the nebular oxygen abundance.

These new nebular abundance measurements also allow us a detailed
comparison with the chemical abundances measured in the neutral gas. 
Measurements of ultraviolet absorption lines obtained from 
{\em FUSE\/} observations have been reported by \cite{heckman01} and
are listed in Table~\ref{table_compare}.  
\cite{heckman01} noted the very low value of N/H in the neutral gas,
but concluded that most of the rest of the measured abundances were
consistent with the nominal metallicity for NGC~1705.
From Table~\ref{table_compare} we see that the N/H abundance in the
neutral gas is, indeed, lower than that in the nebular gas by 
$\sim$~0.7 dex.  
However, the O/H abundance in the neutral gas is also significantly
lower by $\sim$~0.8 dex.  
The lower O/H abundance in the neutral gas is similar to the results
of other observations of dwarf galaxies 
(e.g., I~Zw~18 - \citealp{aloisi03}, NGC~625 - \citealp{cs04}).  
Note that the \ion{O}{1} $\lambda$ 1039.2~\AA\ line can be at or near 
saturation for typical \hi\ column densities (e.g., I~Zw~18 - 
\citealp{lecavelier04}) and this implies that the O/H measurement
may be better interpreted as a lower limit.  
Nonetheless, the N/H abundances have generally been thought to be
secure. 

That the N/O abundance ratio in the neutral gas is essentially
identical to that in ionized gas, yet the N/H and O/H are depressed
by roughly a factor of five, could be taken as support for the 
premise that the ultraviolet absorption lines are probing a ``halo''
of neutral gas with a lower metallicity than the main disk of gas
in NGC~1705.  
Note that the geometry of NGC~1705 may be most favorable for this
interpretation.  
Although we appear to be viewing the disk of NGC~1705 face-on 
\citep{meurer98}, the extinction to the SSC is negligible and the \hi\
column is only $1.6 \times 10^{20}$ atoms cm$^{-2}$ \citep{heckman01}.
This low column density implies that the ultraviolet absorption lines
are not probing the disk gas where column densities are higher.  
This is in contrast to the case of I~Zw~18 where the \hi\ column
density from ultraviolet absorption is more than an order of magnitude 
larger, and therefore almost certainly is probing the main disk gas.
Measuring the metallicity of the gas in the outer parts of dwarf
galaxies remains a challenge and has important implications for the
evolutionary status of dwarf galaxies (cf., \citealp{ks01}).
If a number of dwarf galaxies with characteristics similar to NGC~1705
show this trend for lower metallicity in their halo gas, then future
studies of the chemical evolution of dwarf galaxies should consider
modeling the metallicities with radial gradients (as is the case for
spiral galaxies).

\subsubsection{Spatial Variations}

The spectra of \hii\ regions in spiral galaxies have shown that
oxygen abundances in spiral galaxies decrease with increasing
galactocentric radii 
(e.g., \citealp{mrs85,zkh94,afflerbach96,kbg03}).
However, in dwarf galaxies where there is often only a single \hii\
region, it is assumed that nebular oxygen abundances are
representative of the interstellar medium metallicity for the entire
galaxy.
This appears to be true from extensive spectroscopic mappings of \hii\
regions in dwarf and starburst galaxies; e.g.,
NGC~1569 \citep{devost97,ks97}, NGC~2366 \citep{roy96}, 
NGC~4214 \citep{ks96}, NGC~5253 \citep{chip97}, 
NGC~6822 (\citealp{pes80}; Lee, Skillman, \& Venn, in preparation).
\cite{ks97} discussed scenarios to explain flat chemical abundance
profiles seen on spatial scales between $\ga$ 10~pc and 
$\la$ 1~kpc in low-mass galaxies.
A possible explanation is that freshly produced chemical elements from
are dispersed or transported very quickly to all regions of the galaxy
and are well mixed very quickly as well on small scales within the
interstellar medium. 
However, the timescales required for both dispersal and mixing
must be less than 10$^7$ yr.
The most plausible scenario is that newly synthesized heavy elements
are not yet mixed with the surrounding gas in the interstellar medium,
and that these fresh metals reside in relatively hot $10^6$~K gas
or in cold molecular gas.
Both of these phases are at present difficult to detect. 
%

The metallicity for the SSC was found to be about one-half solar, or
12$+$log(O/H) $\sim$ 8.4, but this appears only to be a rough estimate
\citep{vazquez04}.
The presence of massive star clusters should be able to provide
substantial localized chemical enrichment (see \S~5 in \citealp{ks97}
and references therein). 
\hii\ regions where \othreea\ measurements were presented above are
distributed in a radius of about 10\arcsec\ from the SSC.
From the present data, direct oxygen abundances are in the range
between 12$+$log(O/H) = 8.17 and 8.29. 
Individual variations (0.04 to 0.08 dex) for \hii\ region abundances
about the weighted mean are comparable with the 0.05~dex uncertainty
computed for the mean.
This suggests a lack of spatial variations in oxygen abundances
for \hii\ regions where \othreea\ was measured.
This is consistent with similar conclusions for other dwarf
galaxies; for example, in the post-starburst galaxy NGC~1569 
\citep{ks97}. 

We note with interest that {\em Chandra\/} observations of NGC~1569 by 
\cite{martin02} have revealed the existence of metals-enriched winds
in the hot gas phase, and that the wind has carried nearly all of the
metals ejected by the recent starburst.
As \cite{ks97} suggest, we may be seeing good dispersal of newly
synthesized metals from the recent starburst into the interstellar
medium, but not necessarily good mixing of the metals in the various
gas phases within dwarf galaxies. 
Instantaneous recycling is questionable as an appropriate assumption
in simple models of chemical evolution; see also the discussion in
\cite{ks97}.
It is puzzling why we happen to be seeing spatial homogeneity
in nebular oxygen abundances at this very time, especially if 
the metallicity of the most recent starburst in the SSC is indeed
confirmed to be higher than that of the nebular phase.
It will be interesting to see if {\em Chandra\/} observations can shed
any light on the metallicity of the hot gas phase, although NGC~1705
may be too faint in X-ray luminosity to place useful constraints.

\subsection{Evolution of NGC 1705}

From the \halpha\ flux reported by \cite{gildepaz03}
(see Table~\ref{table_n1705}), the total \halpha\ luminosity is
$L$(\halpha) = $7.9 \times 10^{39}$ ergs~s$^{-1}$. 
Using the formulation by \cite{ktc94}, the \halpha\ luminosity is
converted to the total current star formation rate (SFR) as 
\begin{equation}
{\rm SFR} = 
\frac{L(\halpha)}{1.26 \times 10^{41}\;{\rm ergs\;s}^{-1}} \;
M_{\odot}\;{\rm yr}^{-1}.
\label{eqn_sfr}
\end{equation}
This gives an SFR equal to 0.06 \msun\ yr$^{-1}$.
The star formation rate per unit luminosity is given by 
SFR/$L_B$ = 
$2.3 \times 10^{-10}$ \msun\ yr$^{-1}$ \lsun$^{-1}$.
The formation timescale is thus $L_B$/SFR = 4.3~Gyr.
The gas depletion timescale is given by the ratio of the
total gas mass (1.3 $M_{\rm HI}$ to account for helium) to
the SFR, and is equal to 2.1~Gyr.
The consumption and formation timescales are much lower than normal
irregulars (e.g., \citealp{scm03a}), but this is not surprising, if
the burst in NGC~1705 is a large contributor to the luminosity in $B$. 

The metallicity-luminosity diagram has been known as being
representative of a relationship between metallicity and stellar mass
for dwarf irregular galaxies (e.g.,
\citealp{skh89,rm95,lee03field,scm03b}).
The measured oxygen abundance and $B$ luminosity for NGC~1705 are
consistent with the relation defined by nearby dwarf irregular
galaxies (Fig.~\ref{fig_zlum}).
\cite{lee03virgo} noted that oxygen abundances for dwarf irregulars in
the Virgo Cluster were comparable to a local sample of dwarf
irregulars at a given luminosity.
The metallicity-luminosity diagram for dwarf irregulars appears
relatively unchanged by the cluster environment, although it is
thought that dwarf galaxies should be affected by both external 
(e.g., tidal and/or ram-pressure stripping) and internal (e.g.,
blowout) mechanisms.
NGC~1705 is a relatively isolated galaxy, and Fig.~\ref{fig_zlum}
shows that the {\em current\/} state of the galaxy in the
metallicity-luminosity diagram appears relatively unaffected by the
internal trauma caused by the recent starburst.

However, if the data point for NGC~1705 is indeed above and to the
left of the best-fit line shown in the Figure, possible scenarios to
account for the apparent offset include constant star formation rate
(to boost the metallicity at a given luminosity), pure fading (to
decrease the luminosity at a given metallicity), or both 
(e.g., Fig.~9 in \citealp{chip03}).
Observations of the SSC have shown that star formation ceased
$\approx$ 10~Myr ago, but any fading in 10~Myr is negligible.
For example, timescales of order 1~Gyr are required for passive fading
by about 1~mag (e.g., \citealp{bothun82}).
\cite{annibali03} have shown that continuous star formation with a
variable rate can explain the age mixture of stellar populations
observed in their {\em HST\/} photometry.
If we assume a roughly constant (integrated over the last $\approx$
1~Gyr) rate of star formation, it is possible that star formation has
supplied additional metals to explain how the nebular oxygen abundance
for NGC~1705 may be higher than expected compared to local dwarfs at
similar luminosity. 
On the other hand, there may be no offset and, because of the
relatively small number of galaxies shown in the plot, this may simply
be due to scatter in the metallicity-luminosity diagram.

Dwarf galaxies can experience large outflows, which has been 
observed in NGC~1705 \citep{meurer92,marlowe95,heckman01}.
We examine the effects on the current gas fraction, which is a key
parameter in models of chemical evolution (e.g., \citealp{pagel97}).
For the simple closed-box model, the fraction of gas mass in the
form of metals (in this case, oxygen), $Z_O$, is given by
\begin{equation}
Z_O = y_O \, \ln (1/\mu),
\end{equation}
where $y_O$ is the mass fraction of metals in the form of oxygen,
and $\mu$ is the gas fraction, which is the ratio of gas mass to
the total mass in gas and stars.
The yield is usually referred as the ``effective yield'' if
the assumptions in the closed-box model are not appropriate
(e.g., \citealp{edmunds90,garnett02}).
The total gas mass is usually the \hi\ gas mass multiplied by 
a constant (1.3) to account for helium; the molecular and ionized gas
masses are assumed to be small.
Baryonic gas fractions also require knowledge of stellar masses.
A constant stellar mass to light ratio is usually applied, although
there may be some variation from galaxy to galaxy; see, e.g., 
\cite{belldejong01} and \cite{lee03field}.
We consider here the gas to light ratio, $M_{\rm HI}/L_B$, as a proxy
for the gas fraction.
The advantages of using the gas-to-light ratio are:
the ratio is constructed from purely observed quantities, and
the ratio is independent of distance.

Figure~\ref{fig_zmh1lb} shows a plot of oxygen abundance versus
the \hi\ gas to $B$ light ratio for dwarf galaxies.
It is clear from this diagram that there is a rough correlation
among dwarf galaxies where oxygen abundances increase with
decreasing gas to light ratios, as expected for normal astration.
Local dwarfs, Sculptor group dwarfs, a number of more distant dwarfs
in the field, and the nearby gas-rich dwarf DDO~154 are all plotted
for comparison \citep{chip97,vanzee97,ks01,lee03field,scm03b}. 
\cite{scm03b} already noted that a number of isolated dwarfs in the
field and DDO~154 are best fit by a model with effective yield by
mass equal to $4.9 \times 10^{-3}$.
This corresponds to about 90\% of the solar value.
At low oxygen abundance, a number of local dwarfs appear to have lower
than expected $M_{\rm HI}/L_B$ values at a given metallicity.
The bottom (dashed) curve is an arbitrary fit to the lower envelope of
the locus for nearby dwarfs; the effective yield in this case
is about 12\% of the solar value.
The lower gas-to-light ratios could arise from (a) lower gravitational
potentials, which are unable to retain metals, and/or (b) the effect
of a recent burst which contributes to the luminosity in $B$.

The ``current'' state of NGC~1705 puts this galaxy
in similar phase space with the locus of local dwarfs at the upper
range of oxygen abundances.
Adopting a stellar mass to light ratio equal to 1.2, 
the model which fits NGC~1705 has an effective yield equal
to $1.6 \times 10^{-3}$, or about 30\% of the solar value.
Comparing to the sample of field dwarf irregulars with near-solar
yields at a similar oxygen abundance, it is possible that NGC~1705
has experienced previous gas loss to reduce the effective yield from
near solar to one-third solar.
\cite{heckman01} suggested a scenario that most of the \hi\ gas would
be retained in the galaxy, but most of the hot gas driving the outflow
would escape the galaxy.
Mass loss (with the depletion timescale as a rough indicator) will
eventually convert this galaxy into a gas-poor dwarf with an enhanced
stellar core (i.e., the SSC).

One must be cautious about overinterpreting the results.
Although there is significant scatter in Fig.~\ref{fig_zmh1lb},
one might ask about the significance of the positions of IC~5152,
NGC~625, NGC~1569, and NGC~5253 with respect to NGC~1705.
All of the galaxies shown could simply be a part of the overall
distribution of galaxies with similar abundances and a wide range in
$M_{\rm HI}/L_B$. 
Optical colors for the galaxies shown are similar with a range of 
$B-V$ from about 0.4 to 0.8; indeed, it is very difficult to spot a 
real trend or to pick out ``outliers'' on the basis of their color.
For the five galaxies, their \hi\ masses are all roughly the same
(between 1.0 and 1.3 $\times 10^8$ \msun), and their total $B$
luminosities, log $L_B$, are in the range between 8.4 and 9.1.

Could the total blue luminosity be biased by the recent burst 
of star formation (i.e., SSC)?
The optical luminosity of the SSC is 
$M_{{\rm cl},B} = -13.6$ \citep{sm01}.
The relative contribution of $B$ luminosity by the SSC to the total
galaxy luminosity is about 16\%, which is in contrast with the
ultraviolet where the SSC contributes $\sim$ 50\% of the total
emission \citep{meurer95,hl97}. 
Because NGC~1705 is close to face-on and the line of sight to the SSC
is not extincted, one expects that other starburst galaxies with
edge-on inclinations should have smaller relative contributions by
their respective star clusters to the total optical $B$ luminosities.
We see that the luminosity contribution by the recent starburst 
to the total optical light is not negligible.
Gathering a sample of galaxies with infrared fluxes would be
worthwhile 
(1) to judge properly the underlying stellar mass, which has been shown
to be a better indicator of evolution for star-forming galaxies than
luminosity \citep{tremonti04},
and 
(2) to reduce the scatter caused by recent bursts of star formation.
In particular, a comparison between relatively quiescent dwarf
irregulars with dwarf starburst galaxies would be illuminating.

\section{Conclusions}		
\label{sec_concl}		

Optical long-slit spectra have been obtained for 16 \hii\ regions in
NGC~1705. 
\othreea\ is detected in five \hii\ regions, and the resulting oxygen
abundances are between 12$+$log(O/H) = 8.17 and 8.29.
We adopt a mean value of 12$+$log(O/H) = $8.21 \pm 0.05$, 
corresponding to [O/H] = $-$0.45, or 35\% of the solar value.
There is no significant spatial variation of \othreea\ abundances
($\la$ 0.10~dex) in \hii\ regions distributed in a radius of $\approx$
10\arcsec\ from the super star cluster.
Bright-line calibrations are used to derive oxygen abundances
in the remaining \hii\ regions without \othreea\ measurements; 
bright-line oxygen abundances are accurate only to within 0.2~dex.
The mean argon- and nitrogen-to-oxygen abundance ratios are consistent
with average values for blue compact dwarf galaxies and other
samples of dwarf irregular galaxies.
The mean neon-to-oxygen ratio for NGC~1705 is about a factor of two
larger than the value for blue compact dwarf galaxies, although
the present value is still in the range of values for \hii\ galaxies.
The nitrogen-to-oxygen ratio in the ionized \hii\ gas is in agreement
with the value for neutral \hi, even though the metallicity of the
neutral gas may be as much as six times lower than that of the 
ionized gas.
From observed Balmer flux ratios, derived values of the extinction in
$V$, $A_V$, are in the range zero to 0.9~mag along various lines of
sight to \hii\ regions in the galaxy.
One may need to take into account possible spatial variations in
extinction to correct properly the photometry of resolved stars, which
can affect the derived history of star formation.
The position of NGC~1705 on metallicity-$B$ luminosity and
metallicity-gas fraction plots shows that the adopted oxygen
abundance is comparable to Local Group dwarf irregulars at
comparable luminosity and gas fraction.
In combination with existing observations, simple chemical evolution
models suggest that NGC~1705 may be quickly evolving
into a gas-poor dwarf galaxy.

\begin{acknowledgements}	

We thank the anonymous referee for comments which improved the
presentation of this paper. 
We thank Alessandra Aloisi for suggesting this project and providing
valuable comments, 
Ken Sembach and John Cannon for additional comments,
and
Tim Heckman, Gerhardt Meurer, and Dave Strickland 
for helpful communications.
H. L. also thanks Lisa Germany and the staff at ESO La Silla for
their help with the observations.
H. L. and E. D. S. acknowledge partial support from a NASA LTSARP grant
NAG~5--9221 and the University of Minnesota.
This research has made use of NASA's Astrophysics Data System, and
of the NASA/IPAC Extragalactic Database (NED), which is operated by 
the Jet Propulsion Laboratory, California Institute of Technology, 
under contract with the National Aeronautics and Space Administration. 
\end{acknowledgements}

Facilities: \facility{ESO(EFOSC2)}

\clearpage	

\begin{figure}
\figurenum{1}
\caption{
Long-slit orientations in NGC~1705 using the \halpha\ image adapted
from \citet*[][their Fig.~3]{gildepaz03}.
Black objects on the image indicate bright sources.
The field shown in the left frame is $1\farcm56 \times 1\farcm70$.
The contrast is adjusted to highlight both bright and faint \halpha\
emission.
The central $0\farcm78 \times 0\farcm85$ is indicated as a box
and is shown in the right frame. 
\hii\ regions where \othreea\ is detected are labelled.
In both frames, north and east are to the top and the left,
respectively.
The position of the super star cluster (SSC) is indicated
\citep{mmt85,meurer89}; the SSC has been blanked out of the 
\halpha\ image. 
}
\label{fig_n1705ha}
\end{figure}


\begin{figure}
\figurenum{2}
\caption{
Two-minute acquisition image in \halpha\ with EFOSC2 for 
the slit A configuration.
Black objects on the image indicate bright sources.
The separation between the solid lines corresponds approximately
to the 1\farcs5 slit width projected on the sky.
Labeled are \hii\ regions for which spectra were obtained.
In the right panel, the image is rescaled to emphasize \hii\ regions
A1, A2, and A4. 
}
\label{fig_slita}
\end{figure}


\begin{figure}
\figurenum{3}
\caption{
Two-minute acquisition image in \halpha\ for the slit B
configuration.
The aperture for \hii\ region B4 encompassed \hii\ regions B3 and B5.
\hii\ region B8 corresponds to the bright feature in an \halpha\
arc to the northeast; this is also seen as a bright clump in the arc
``A7'' in \cite{meurer92}; see also \cite{gildepaz03}.
In the right panel, the image is rescaled to emphasize \hii\ regions
B1, B7, and B8. 
See Fig.~\ref{fig_slita} for additional comments.
}
\label{fig_slitb}
\end{figure}


\begin{figure}
\figurenum{4}
\caption{
Two-minute acquisition image in \halpha\ for the slit C
configuration. 
The aperture for \hii\ region C2 encompassed \hii\ regions C1 and C3.
See Fig.~\ref{fig_slita} for additional comments.
}
\label{fig_slitc}
\end{figure}


\begin{figure}
\figurenum{5}
\caption{
Emission-line spectra between 3600 and 7300~\AA.
The observed flux per unit wavelength is plotted versus wavelength.
The bottom panel in each part is displayed to highlight fainter
emission lines, especially ${\rm [O\;III]}\lambda$ 4363 as indicated.
(a) \hii\ region A3.
(b) \hii\ region B3.
(c) \hii\ region B6.
(d) \hii\ region C6.
}
\label{fig_specall}
\end{figure}








\begin{figure}
\figurenum{6}
\caption{
Spectrum of aperture C5, which is the super star cluster (SSC) 
first discussed by \cite{mmt85} and \cite{meurer89}.
The top panel shows the spectrum from 3600 to 7300~\AA; note
the weak \othree\ and \halpha\ emission.
The middle and bottom panels show the spectrum from
3600 to 5500~\AA, and from 5500 to 7300~\AA, respectively.
}
\label{fig_ssc}
\end{figure}


\begin{figure}
\figurenum{7}
\caption{
Monte Carlo simulations of solutions for the reddening, $c(\hbeta)$,
and the underlying Balmer absorption with equivalent width, 
EW$_{\rm abs}$, from hydrogen Balmer flux ratios.
Dotted lines mark zero values for each quantity.
The results here are shown for the \hii\ region A3.
Each small point is a solution derived from a different realization
of the same input spectrum.
The large filled circle with error bars shows the mean result with
$1\sigma$ errors derived from the dispersion in the solutions.
}
\label{fig_monte}
\end{figure}


\begin{figure}
\figurenum{8}
\caption{
Spectrum of \hii\ region B3 in the region between 
4300 and 4800~\AA.  
Highlighted are emission lines \hgamma, [\ion{O}{3}]$\lambda$ 4363, 
\ion{He}{1}$\lambda$ 4471, and \ion{He}{2}$\lambda$ 4686.
}
\label{fig_he2}
\end{figure}


\begin{figure}
\figurenum{9}
\caption{
Difference in oxygen abundance from various methods
versus log~$O_{32}$ (top panel), and 
versus log~$R_{23}$ (bottom panel).
Each point represents an \hii\ region.
``Direct'' denotes oxygen abundances derived from \othreea\
measurements, ``M91'' denotes oxygen abundances derived using
the bright-line method by \cite{mcgaugh91}, and ``P00'' denotes
oxygen abundances derived using the bright-line method 
by \cite{pilyugin00}.
Vertical dotted lines in both panels mark zero differences in 
oxygen abundance.
In the bottom panel, we list in the legend the dispersion in the
differences for each method used to derive oxygen abundances.
These plots show that oxygen abundances derived with bright-line
methods are accurate only to within $\approx$ 0.2~dex.
}
\label{fig_oxydiff}
\end{figure}


\begin{figure}
\figurenum{10}
\caption{
Oxygen abundance versus absolute magnitude in $B$ for 
dwarf irregular galaxies.
Only galaxies with \othreea\ detections are shown.
An open star marks the location of NGC~1705.
A sample of local dwarf irregular galaxies is marked
by solid circles and the solid line is a fit to these galaxies
\citep{lee03field,chip97}. 
The ``Y'' symbols are dwarf irregulars from the Sculptor
group \citep{scm03a,scm03b}.
Open triangles indicate a sample of dwarf irregulars in the field
\citep{vanzee97}.
DDO~154 \citep{vanzee97,ks01} is marked by an open square. 
For its luminosity, the oxygen abundance of NGC~1705 is comparable
to other dwarf irregulars.
}
\label{fig_zlum}
\end{figure}


\begin{figure}
\figurenum{11}
\caption{
Oxygen abundance versus \hi\ gas-to-light ratio for dwarf 
irregular galaxies.
The symbols are the same as in Fig.~\ref{fig_zlum}.
Chemical evolution models shown are as follows: 
dotted line - model to fit DDO~154 \citep{ks01};
thick solid line - an approximate fit to NGC~1705;
dashed line - {\em ad hoc\/} lower envelope for the locus defined by
dwarfs in this plot.
A constant stellar mass to light ratio, $M_{\ast}/L_B$ = 1.2, was
adopted for all models.  
The effective oxygen yield by mass, $y_{\rm O}$, for each model is 
also given.
Although NGC~1705 is known to be isolated, the galaxy has a
gas-to-light ratio comparable with values for other dwarf irregulars
in group environments at a given oxygen abundance. 
}
\label{fig_zmh1lb}
\end{figure}

\clearpage	

\begin{table}
\small 
\tablenum{1}
\begin{center}
\renewcommand{\arraystretch}{1.1} 
\caption{
Basic data for NGC~1705.
\vspace*{3mm}
\label{table_n1705}
}
\begin{tabular}{cccc}
\tableline \tableline
& Property & Value & References \\
\tableline
(1) & Type & amorphous; BCD,N & 1, 2 \\
(2) & $F_{21}$ (Jy km s$^{-1}$) & 16.6 & 2 \\
(3) & $D$ (Mpc) & $5.1 \pm 0.6$ & 3 \\ 
(4) & Scale (pc arcsec$^{-1}$) & 25 & 4 \\
(5) & $v_{\odot}$ (km s$^{-1}$) & $628 \pm 9$ & 5 \\
(6) & $v_{\rm rot}$ (km s$^{-1}$) & 62 & 5 \\
(7) & $r_{{\rm exp},B}$ (\arcsec) & 13.6 & 6 \\
(8) & $\mu_{0,B}$ (mag arcsec$^{-2}$) & 20.86 & 6 \\
(9) & $B$ (mag) & $13.09 \pm 0.01$ & 7 \\ 
(10) & $B-R$ (mag) & $0.90 \pm 0.06$ & 7 \\
(11) & $F(\halpha)$ (ergs s$^{-1}$ cm$^{-2}$) & 
       $(2.53 \pm 0.03) \times 10^{-12}$ & 7 \\
(12) & $L_{\rm UV}$ (ergs s$^{-1}$) & $7.4 \times 10^{41}$ & 8 \\
(13) & $L_{\rm X}$ (ergs s$^{-1}$) & $1.2 \times 10^{38}$ & 9 \\
(14) & $E(B-V)_G$ (mag) & $+$0.035 & 10 \\
(15) & $E(B-V)_{\rm i}$ (mag) & 0.00 & 11 \\
(16) & $M_B$ (mag) & $-15.6$ & 4 \\
(17) & $M_{\rm HI}$ (\msun) & $1.0 \times 10^8$ & 4 \\
(18) & log $M_{\rm HI}/L_B$ & $-0.42$ & 4 \\
(19) & 12$+$log(O/H) & $8.21 \pm 0.05$ & 4 \\
%
%
\tableline
\end{tabular}
\tablecomments{
Properties.
(1) Morphological type.
(2) 21-cm flux integral.
(3) Distance.
(4) Linear to angular scale at this distance.
(5) Systemic heliocentric velocity in the optical.
(6) Maximum \hi\ rotational velocity.
(7) Exponential scale length in $B$.
(8) Uncorrected central surface brightness in $B$.
(9) Apparent total magnitude in $B$.
(10) Measured total $B-R$ color.
(11) Integrated \halpha\ flux, corrected for underlying Balmer
absorption with equivalent width 3~\AA.
(12) Total ultraviolet luminosity.
(13) Total X-ray luminosity (soft).
(14) Reddening from the Galactic foreground.
(15) Reddening intrinsic to NGC~1705.
(16) Absolute magnitude in $B$, corrected for $E(B-V)_G$.
(17) \hi\ gas mass.
(18) Logarithm of the \hi-mass to blue luminosity ratio.
(19) Mean oxygen abundance from \hii\ regions.
}
\tablerefs{
1. \cite{marlowe95};
2. \cite{meurer92};
3. \cite{tosi01};
4. present work;
5. \cite{meurer98};
6. \cite{marlowe97};
7. \cite{gildepaz03};
8. \cite{heckman98};
9. \cite{hensler98};
10. \cite{schlegel98};
11. \cite{calzetti94}.
}
\end{center}
\end{table}


\begin{table}
\footnotesize 
\tablenum{2}
\begin{center}
\renewcommand{\arraystretch}{1.1} 
\caption{
Properties of EFOSC2 spectrograph employed at the ESO La Silla
3.6-m telescope.
\vspace*{3mm}
\label{table_obsprops}
}
\begin{tabular}{cc}
\tableline \tableline
\multicolumn{2}{c}{{\sf Loral CCD (\#40)}} \\ 
\tableline
Total area & 2048 pix $\times$ 2048 pix \\
Field of view & 5.2 arcmin $\times$ 5.2 arcmin \\
Pixel size & 15 $\mu$m \\
Image scale & 0.16 arcsec pixel$^{-1}$ \\
Gain & 1.3 $e^-$ ADU$^{-1}$ \\ 
Read-noise (rms) & 9 $e^-$ \\ 
\tableline
\multicolumn{2}{c}{{\sf Long slit}} \\ \tableline
Length & $\simeq 5$ arcmin \\
Width & 1.5 arcsec \\
\tableline
\multicolumn{2}{c}{\sf Grating \#11} \\ \tableline
Groove density & 300 lines mm$^{-1}$  \\
Blaze $\lambda$ (1st order) & 4000 \AA \\
Dispersion & 2.04 \AA\ pixel$^{-1}$ \\
Effective $\lambda$ range & 3380--7520 \AA \\ 
%
\tableline
\end{tabular}
\end{center}
\end{table}


\begin{deluxetable}{cccccccccc}
\tabletypesize{\footnotesize} 
\tablenum{3}
\renewcommand{\arraystretch}{1.1} 
\tablecolumns{10}
\tablewidth{0pt}
\tablecaption{
Log of Observations.
\label{table_obslog}
}
\tablehead{
& \colhead{Date} & 
\colhead{$\alpha$ (J2000)} & \colhead{$\delta$ (J2000)} & PA & 
& \colhead{$t_{\rm total}$} & & & \colhead{RMS} \\ 
\colhead{Slit} & \colhead{(UT 2003)} & 
\colhead{$(^h\;^m\;^s)$} & \colhead{$(\degr \; \arcmin \; \arcsec)$} &
\colhead{($\degr$)} & \colhead{$N_{\rm exp}$} & \colhead{(s)} & 
\colhead{$\langle X \rangle$} & \colhead{[\ion{O}{3}]$\lambda$ 4363} & 
\colhead{(mag)} \\
\colhead{(1)} & \colhead{(2)} & \colhead{(3)} & \colhead{(4)} &
\colhead{(5)} & \colhead{(6)} & \colhead{(7)} & \colhead{(8)} &
\colhead{(9)} & \colhead{(10)} 
}
\startdata
A & 27 Aug & $04\;54\;17.0$ & $-53\;21\;34.0$ & 275\fdg0 &
$3 \times 1200$ & 3600 & 1.21 & A3 & 0.029 \\
B & 28 Aug & $04\;54\;15.4$ & $-53\;21\;39.4$ & 358\fdg1 &
$1 \times 1200$ & 1200 & 1.18 & B3, B4, B6 & 0.034 \\
C & 28 Aug & $04\;54\;16.2$ & $-53\;21\;17.7$ & 243\fdg2 &
$1 \times 1200$ & 1200 & 1.18 & C6 & 0.034 \\
\enddata
\tablecomments{
Col.~(1): Slit orientation, as shown in Fig.~\ref{fig_n1705ha}.
Col.~(2): Date.
Cols.~(3) and (4): Center of the long-slit in 
right ascension and declination (Epoch J2000).
Col.~(5): Position angle, north through east.
Col.~(6): Number of exposures obtained and the length of each
exposure in seconds. 
Col.~(7): Total exposure time.
Col.~(8): Mean effective airmass.
Col.~(9): \hii\ regions where \othreea\ detected; see
Table~\ref{table_h2reglist}.
Col.~(10): Relative root--mean--square error in the sensitivity
function obtained from observations of standard stars.
}
\end{deluxetable}


\begin{table}
\tablenum{4}
\begin{center}
\renewcommand{\arraystretch}{1.1} 
\caption{
H~II regions in NGC~1705.
\vspace*{3mm}
\label{table_h2reglist}
}
\begin{tabular}{ccc}
\tableline \tableline
\hii & & \\ 
Region & MMT85 & MFDC92 \\
(1) & (2) & (3) \\
\tableline
%
%
A1 & \nodata & \nodata \\
A2 & B & H2 \\
A3 & \nodata & H1 \\
A4 & \nodata & \nodata \\
B1 & C & H3 \\
B2 & \nodata & \nodata \\
B3 & \nodata & \nodata \\
B4$\,$\tablenotemark{a} & \nodata & \nodata \\
B5 & \nodata & \nodata \\
B6$\,$\tablenotemark{b} & D & H4 \\
B7 & \nodata & \nodata \\
B8 & \nodata & \nodata \\
C1 & \nodata & H5 \\
C2$\,$\tablenotemark{c} & \nodata & \nodata \\
C3$\,$\tablenotemark{b} & D & H4 \\
C4 & \nodata & \nodata \\
C5$\,$\tablenotemark{d} & A & N \\
C6$\,$\tablenotemark{e} & \nodata & \nodata \\
\tableline
\end{tabular}
\tablenotetext{a}{
The pixels defining the extraction aperture for B4 encompass the \hii\
regions B3 and B5.
}
\tablenotetext{b}{
B6 and C3 are the same \hii\ region.
}
\tablenotetext{c}{
The pixels defining the extraction aperture for C2 encompass the \hii\
regions C1 and C3.
}
\tablenotetext{d}{
Super star cluster (SSC), or NGC~1705-1.
The spectrum of the SSC shown in Fig.~\ref{fig_ssc} exhibits weak
\othree\ and \halpha\ emission, also seen previously by \cite{mmt85}.
This spectrum is not included in the present analysis.  
}
\tablenotetext{e}{
The pixels defining the extraction aperture for C6 encompasses light
from \hii\ regions A3 and A4.
}
\tablecomments{
Col.~(1): \hii\ regions identified in Figs.~\ref{fig_slita}
to \ref{fig_slitc}.
Cols.~(2) and (3): Identifications with previous work --
MMT85: \citet[][their Fig.~2]{mmt85}, 
MFDC92: \citet[][their Table~5 and Fig.~12]{meurer92}.
}
\end{center}
\end{table}


\begin{table}
\scriptsize 
\tablenum{5a}
\begin{center}
\renewcommand{\arraystretch}{0.9} 
\caption{
Line ratios and properties for \hii\ regions A1 to A4 inclusive.
\vspace*{3mm}
\label{table_data1}
}
\begin{tabular}{rccccccc}
\tableline \tableline
& & \multicolumn{2}{c}{A1} & 
\multicolumn{2}{c}{A2} & 
\multicolumn{2}{c}{A3} \\
\multicolumn{1}{c}{Wavelength (\AA)} &
\multicolumn{1}{c}{$f(\lambda)$} &
\multicolumn{1}{c}{$F$} & \multicolumn{1}{c}{$I$} &
\multicolumn{1}{c}{$F$} & \multicolumn{1}{c}{$I$} &
\multicolumn{1}{c}{$F$} & \multicolumn{1}{c}{$I$} \\
\tableline
$[\rm{O\;II}]\;3727$ & $+0.325$ &
	  $324.5 \pm 7.9$ & $374 \pm 34$ &
	  $242.9 \pm 6.5$ & $337 \pm 62$ &
	  $221.5 \pm 4.4$ & $254 \pm 12$
\\
$[{\rm Ne\;III}]\;3869$ & $+0.294$ &
	  $36.6 \pm 6.0$ & $41.0 \pm 9.0$ &
	  $50.9 \pm 3.4$ & $66 \pm 14$ &
	  $43.0 \pm 1.2$ & $48.6 \pm 2.4$
\\
${\rm H}8 + {\rm He\;I}\;3889$ & $+0.289$ &
	  $15.0 \pm 6.2$ & $20 \pm 11$ &
	  \nodata & \nodata &
	  $16.0 \pm 1.1$ & $24.3 \pm 3.2$
\\
${\rm H}\epsilon + {\rm He\;I}\;3970$\tablenotemark{a} & $+0.269$ &
	  $27.5 \pm 3.1$ & $32.9 \pm 6.0$ &
	  \nodata & \nodata &
	  $20.88 \pm 0.90$ & $28.3 \pm 1.5$
\\
${\rm He\;I}\;4027$ & $+0.253$ &
	  \nodata & \nodata & 
	  \nodata & \nodata & 
	  $1.84 \pm 0.68$ & $2.04 \pm 0.76$
\\
${\rm H}\delta\;4101$ & $+0.232$ &
	  $19.0 \pm 3.5$ & $23.7 \pm 6.3$ &
	  \nodata & \nodata &
	  $19.7 \pm 1.0$ & $26.1 \pm 1.4$
\\
${\rm H}\gamma\;4340$ & $+0.158$ &
	  $39.9 \pm 3.3$ & $43.8 \pm 6.5$ &
	  $27.8 \pm 2.4$ & $40 \pm 10$ &
	  $43.9 \pm 1.4$ & $49.8 \pm 1.8$ 
\\
$[{\rm O\;III}]\;4363$ & $+0.151$ &
	  $< 9.97$ & $< 10.3$ &
	  \nodata & \nodata &
	  $3.9 \pm 1.1$ & $4.1 \pm 1.2$
\\
${\rm He\;I}\;4471$ & $+0.116$ &
	  \nodata & \nodata &
	  \nodata & \nodata &
	  $3.48 \pm 0.55$ & $3.60 \pm 0.57$
\\
${\rm H}\beta\;4861$ & \phs0.000 &
	  $100.0 \pm 4.2$ & $100.0 \pm 5.3$ &
	  $100 \pm 13$ & $100 \pm 14$ &
	  $100.0 \pm 3.4$ & $100.0 \pm 3.3$
\\
$[{\rm O\;III}]\;4959$ & $-0.026$ &
	  $109.1 \pm 8.6$ & $105 \pm 14$ &
	  $185 \pm 13$ & $166 \pm 37$ &
	  $153 \pm 11$ & $147 \pm 11$
\\
$[{\rm O\;III}]\;5007$ & $-0.038$ &
	  $312 \pm 11$ & $300 \pm 30$ &
	  $547 \pm 17$ & $487 \pm 91$ &
	  $444 \pm 15$ & $425 \pm 14$ 
\\
${\rm He\;I}\;5876$ & $-0.204$ &
	  $10.2 \pm 4.0$ & $9.1 \pm 4.1$ &
	  $15.1 \pm 2.6$ & $11.4 \pm 3.7$ &
	  $11.72 \pm 0.81$ & $10.3 \pm 0.77$
\\
${\rm H}\alpha\;6563$ & $-0.299$ &
	  $329 \pm 11$ & $286 \pm 28$ &
	  $409 \pm 13$ & $286 \pm 54$ &
	  $346.0 \pm 9.7$ & $291 \pm 15$
\\
$[{\rm N\;II}]\;6583$ & $-0.302$ &
	  $12.2 \pm 9.1$ & $10.5 \pm 8.5$ &
	  \nodata & \nodata &
	  $11.1 \pm 8.1$ & $9.3 \pm 6.8$
\\
${\rm He\;I}\;6678$ & $-0.314$ &
	  \nodata & \nodata &
	  \nodata & \nodata &
	  $5.1 \pm 1.3$ & $4.2 \pm 1.1$
\\
$[{\rm S\;II}]\;6716,6731$ & $-0.320$ &
	  $69.5 \pm 8.5$ & $60 \pm 11$ &
	  $38.8 \pm 3.9$ & $26.2 \pm 6.6$ &
	  $40.8 \pm 1.7$ & $33.8 \pm 2.1$ 
\\
${\rm He\;I}\;7065$ & $-0.366$ &
	  \nodata & \nodata &
	  \nodata & \nodata &
	  $4.28 \pm 0.70$ & $3.46 \pm 0.59$
\\
$[{\rm Ar\;III}]\;7136$ & $-0.375$ &
	  \nodata & \nodata &
	  \nodata & \nodata &
	  $9.15 \pm 0.96$ & $7.3 \pm 1.1$
\\[1mm]
\multicolumn{2}{c}{$F(\hbeta)$ (ergs s$^{-1}$ cm$^{-2}$)} & 
	  \multicolumn{2}{c}{$(4.04 \pm 0.17) \times 10^{-16}$} &
	  \multicolumn{2}{c}{$(7.53 \pm 0.96) \times 10^{-16}$} &
	  \multicolumn{2}{c}{$(3.61 \pm 0.12) \times 10^{-15}$} 
\\
\multicolumn{2}{c}{EW$_{\rm e}$(\hbeta) (\AA)} &
	  \multicolumn{2}{c}{$76.1 \pm 5.1$} &
	  \multicolumn{2}{c}{$23.4 \pm 3.2$} &
	  \multicolumn{2}{c}{$80.1 \pm 4.5$} 
\\
\multicolumn{2}{c}{Derived $E(B-V)$ (mag)} &
	  \multicolumn{2}{c}{$+0.122 \pm 0.099$} &
	  \multicolumn{2}{c}{$+0.30 \pm 0.19$} &
	  \multicolumn{2}{c}{$+0.172 \pm 0.089$} 
\\
\multicolumn{2}{c}{$c(\hbeta)$} &
	  \multicolumn{2}{c}{\nodata} &
	  \multicolumn{2}{c}{\nodata} &
	  \multicolumn{2}{c}{$+0.221 \pm 0.060$} 
\\
\multicolumn{2}{c}{Adopted $A_V$ (mag)} &
	  \multicolumn{2}{c}{$+0.37$} &
	  \multicolumn{2}{c}{$+0.92$} &
	  \multicolumn{2}{c}{$+0.47$} 
\\
\multicolumn{2}{c}{EW$_{\rm abs}$ (\AA)} &
	  \multicolumn{2}{c}{2} &
	  \multicolumn{2}{c}{2} &
	  \multicolumn{2}{c}{$2.1 \pm 1.1$} 
\\[1mm]
\tableline
& & \multicolumn{2}{c}{A4} \\ 
\multicolumn{1}{c}{Wavelength (\AA)} &
\multicolumn{1}{c}{$f(\lambda)$} &
\multicolumn{1}{c}{$F$} & \multicolumn{1}{c}{$I$} \\
\tableline
$[\rm{O\;II}]\;3727$ & $+0.325$ &
	  $248 \pm 33$ & $243 \pm 73$ \\
${\rm H}\beta\;4861$ & \phs0.000 &
	  $100 \pm 15$ & $100 \pm 16$ \\
$[{\rm O\;III}]\;4959$ & $-0.026$ &
          $155 \pm 14$ & $152 \pm 39$ \\
$[{\rm O\;III}]\;5007$ & $-0.038$ &
	  $390 \pm 21$ & $383 \pm 84$ \\
${\rm H}\alpha\;6563$ & $-0.299$ &
	  $277 \pm 25$ & $276 \pm 72$ 
\\[1mm]
\multicolumn{2}{c}{$F(\hbeta)$ (ergs s$^{-1}$ cm$^{-2}$)} & 
          \multicolumn{2}{c}{$(1.14 \pm 0.17) \times 10^{-15}$} \\
\multicolumn{2}{c}{EW$_{\rm e}$(\hbeta) (\AA)} &
          \multicolumn{2}{c}{$101 \pm 30$} \\
\multicolumn{2}{c}{Derived $E(B-V)$ (mag)} &
          \multicolumn{2}{c}{$-0.038 \pm 0.259$} \\
\multicolumn{2}{c}{$c(\hbeta)$} &
          \multicolumn{2}{c}{\nodata} \\
\multicolumn{2}{c}{Adopted $A_V$ (mag)} &
          \multicolumn{2}{c}{0} \\
\multicolumn{2}{c}{EW$_{\rm abs}$ (\AA)} &
          \multicolumn{2}{c}{2} \\
\tableline
\end{tabular}
\tablenotetext{a}{
Blended with [\ion{Ne}{3}]$\lambda$ 3968.
}
\tablecomments{
Emission lines are listed in \AA.
$F$ is the observed flux ratio with respect to \hbeta.
$I$ is the corrected intensity ratio, corrected for the adopted
reddening listed, and for underlying Balmer absorption.
The uncertainties in the observed line ratios account for the
uncertainties in the fits to the line profiles, the surrounding
continua, and the relative uncertainty in the sensitivity function
listed in Table~\ref{table_obslog}. 
Flux uncertainties in the \hbeta\ reference line are not included.
Uncertainties in the corrected line ratios account for uncertainties
in the specified line and in the \hbeta\ reference line.
The reddening function, $f(\lambda)$, from
Equation~(\ref{eqn_corr}) is given.
Also listed are the observed \hbeta\ flux; the equivalent width of
\hbeta\ in emission, EW$_{\rm e}$(\hbeta); derived values of the
reddenings from SNAP using Equation~(\ref{eqn_corrthesis}). 
Where \othreea\ is measured, simultaneous solutions for the
logarithmic reddening, $c(\hbeta)$, from Equation~(\ref{eqn_corr}) and
the equivalent width of the underlying Balmer absorption,
EW$_{\rm abs}$, are listed.
The adopted value of the extinction in $V$, $A_V$, is listed.
Where \othreea\ is not measured, the equivalent width of the
underlying Balmer absorption was set to 2~\AA.
}
\end{center}
\end{table}


\begin{table}
\tiny 
\tablenum{5b}
\begin{center}
\renewcommand{\arraystretch}{1.} 
\caption{
Line ratios and properties for \hii\ regions B1 to B8 inclusive.
\vspace*{3mm}
\label{table_data2}
}
\begin{tabular}{rccccccc}
\tableline \tableline
& & \multicolumn{2}{c}{B1} & 
\multicolumn{2}{c}{B2} & 
\multicolumn{2}{c}{B3} \\ 
\multicolumn{1}{c}{Wavelength (\AA)} &
\multicolumn{1}{c}{$f(\lambda)$} &
\multicolumn{1}{c}{$F$} & \multicolumn{1}{c}{$I$} &
\multicolumn{1}{c}{$F$} & \multicolumn{1}{c}{$I$} &
\multicolumn{1}{c}{$F$} & \multicolumn{1}{c}{$I$} \\
\tableline
$[\rm{O\;II}]\;3727$ & $+0.325$ &
	  $414 \pm 14$ & $400 \pm 51$ &
	  $462 \pm 13$ & $475 \pm 60$ &
	  $300.7 \pm 6.3$ & $274 \pm 7.1$
\\
$[{\rm Ne\;III}]\;3869$ & $+0.294$ &
	  $48.6 \pm 9.4$ & $47 \pm 13$ &
	  $65.7 \pm 6.5$ & $66 \pm 12$ &
	  $67.0 \pm 2.9$ & $61.0 \pm 2.8$
\\
${\rm H}\epsilon + {\rm He\;I}\;3970$ & $+0.269$ &
	  \nodata & \nodata &
	  \nodata & \nodata &
	  $9.9 \pm 2.2$ & $22.2 \pm 2.0$
\\
${\rm H}\delta\;4101$ & $+0.232$ &
	  \nodata & \nodata &
	  \nodata & \nodata &
	  $14.4 \pm 2.0$ & $25.6 \pm 1.8$
\\
${\rm H}\gamma\;4340$ & $+0.158$ &
	  $62.2 \pm 6.7$ & $64 \pm 13$ &
	  $40.9 \pm 3.7$ & $50 \pm 10$ &
	  $39.5 \pm 2.1$ & $46.8 \pm 1.9$
\\
$[{\rm O\;III}]\;4363$ & $+0.151$ &
	  \nodata & \nodata &
	  \nodata & \nodata &
	  $6.1 \pm 1.6$ & $5.6 \pm 1.5$
\\
${\rm He\;I}\;4471$ & $+0.116$ &
	  \nodata & \nodata &
	  \nodata & \nodata &
	  $3.94 \pm 0.94$ & $3.59 \pm 0.86$
\\
${\rm He\;II}\;4686$ & $+0.050$ &
	  \nodata & \nodata &
	  \nodata & \nodata &
	  $5.42 \pm 0.63$ & $4.93 \pm 0.57$ 
\\
${\rm H}\beta\;4861$ & \phs0.000 &
	  $100.0 \pm 6.7$ & $100.0 \pm 7.9$ &
	  $100.0 \pm 6.7$ & $100.0 \pm 8.2$ &
	  $100.0 \pm 4.5$ & $100.0 \pm 4.1$ 
\\
$[{\rm O\;III}]\;4959$ & $-0.026$ &
	  $106.5 \pm 8.8$ & $103 \pm 17$ &
	  $147 \pm 12$ & $132 \pm 22$ &
	  $183 \pm 14$ & $167 \pm 13$
\\
$[{\rm O\;III}]\;5007$ & $-0.038$ &
	  $326 \pm 11$ & $315 \pm 40$ &
	  $412 \pm 15$ & $367 \pm 49$ &
	  $534 \pm 18$ & $486 \pm 16$ 
\\
${\rm He\;I}\;5876$ & $-0.204$ &
	  \nodata & \nodata &
	  \nodata & \nodata &
	  $8.59 \pm 0.94$ & $7.82 \pm 0.86$
\\
${\rm H}\alpha\;6563$ & $-0.299$ &
	  $292 \pm 10$ & $285 \pm 37$ &
	  $340 \pm 11$ & $286 \pm 38$ &
	  $302.5 \pm 9.1$ & $281.3 \pm 9.1$
\\
$[{\rm N\;II}]\;6583$ & $-0.302$ &
	  $4.1 \pm 8.6$ & $3.9 \pm 8.7$ &
	  $14.9 \pm 4.2$ & $12.2 \pm 4.5$ &
	  $10.8 \pm 1.8$ & $9.8 \pm 1.6$
\\
${\rm He\;I}\;6678$ & $-0.314$ &
	  \nodata & \nodata &
	  \nodata & \nodata &
	  $3.1 \pm 1.1$ & $2.8 \pm 1.0$
\\
$[{\rm S\;II}]\;6716,6731$ & $-0.320$ &
	  $50.9 \pm 6.4$ & $49 \pm 10$ &
	  $20.2 \pm 3.6$ & $16.4 \pm 4.4$ &
	  $40.8 \pm 3.0$ & $37.1 \pm 2.8$ 
\\
$[{\rm Ar\;III}]\;7136$ & $-0.375$ &
	  \nodata & \nodata &
	  \nodata & \nodata &
	  $9.8 \pm 1.4$ & $9.2 \pm 1.9$
\\[1mm]
\multicolumn{2}{c}{$F(\hbeta)$ (ergs s$^{-1}$ cm$^{-2}$)} & 
	  \multicolumn{2}{c}{$(9.92 \pm 0.66) \times 10^{-16}$} &
	  \multicolumn{2}{c}{$(1.444 \pm 0.096) \times 10^{-15}$} &
	  \multicolumn{2}{c}{$(5.30 \pm 0.24) \times 10^{-15}$} 
\\
\multicolumn{2}{c}{EW$_{\rm e}$(\hbeta) (\AA)} &
	  \multicolumn{2}{c}{$59.1 \pm 5.5$} &
	  \multicolumn{2}{c}{$18.1 \pm 1.3$} &
	  \multicolumn{2}{c}{$35.5 \pm 1.8$} 
\\
\multicolumn{2}{c}{Derived $E(B-V)$ (mag)} &
	  \multicolumn{2}{c}{$-0.003 \pm 0.130$} &
	  \multicolumn{2}{c}{$+0.096 \pm 0.132$} &
	  \multicolumn{2}{c}{$+0.013 \pm 0.107$} 
\\
\multicolumn{2}{c}{$c(\hbeta)$} &
	  \multicolumn{2}{c}{\nodata} &
	  \multicolumn{2}{c}{\nodata} &
	  \multicolumn{2}{c}{$-0.009 \pm 0.086$} 
\\
\multicolumn{2}{c}{Adopted $A_V$ (mag)} &
	  \multicolumn{2}{c}{0} &
	  \multicolumn{2}{c}{$0.29$} &
	  \multicolumn{2}{c}{0} 
\\
\multicolumn{2}{c}{EW$_{\rm abs}$ (\AA)} &
	  \multicolumn{2}{c}{2} &
	  \multicolumn{2}{c}{2} &
	  \multicolumn{2}{c}{$3.7 \pm 1.2$} 
\\[1mm]
\tableline
& & \multicolumn{2}{c}{B4} & 
\multicolumn{2}{c}{B5} & 
\multicolumn{2}{c}{B6} \\
\multicolumn{1}{c}{Wavelength (\AA)} &
\multicolumn{1}{c}{$f(\lambda)$} &
\multicolumn{1}{c}{$F$} & \multicolumn{1}{c}{$I$} &
\multicolumn{1}{c}{$F$} & \multicolumn{1}{c}{$I$} &
\multicolumn{1}{c}{$F$} & \multicolumn{1}{c}{$I$} \\
\tableline
$[\rm{O\;II}]\;3727$ & $+0.325$ &
	  $345.4 \pm 7.2$ & $312 \pm 8.0$ &
	  $398.8 \pm 8.8$ & $349 \pm 36$ &
	  $484.7 \pm 9.9$ & $436 \pm 19$
\\
$[{\rm Ne\;III}]\;3869$ & $+0.294$ & 
	  $80.4 \pm 4.2$ & $72.7 \pm 3.8$ &
	  $103.2 \pm 7.1$ & $90 \pm 13$ &
	  $77.6 \pm 5.2$ & $69.9 \pm 5.3$
\\
${\rm H}\gamma\;4340$ & $+0.158$ &
	  $37.6 \pm 2.1$ & $45.8 \pm 1.9$ &
	  $33.7 \pm 2.7$ & $45.2 \pm 8.1$ &
	  $39.9 \pm 2.0$ & $46.9 \pm 2.0$
\\
$[{\rm O\;III}]\;4363$ & $+0.151$ &
	  $5.3 \pm 1.6$ & $4.8 \pm 1.5$ &
	  \nodata & \nodata &
	  $6.3 \pm 1.6$ & $5.7 \pm 1.5$
\\
${\rm He\;I}\;4471$ & $+0.116$ &
	  $2.68 \pm 0.92$ & $2.42 \pm 0.83$ &
	  \nodata & \nodata &
	  \nodata & \nodata
\\
${\rm He\;II}\;4686$ & $+0.050$ &
	  $3.96 \pm 0.57$ & $3.58 \pm 0.52$ &
	  \nodata & \nodata &
	  \nodata & \nodata 
\\
${\rm H}\beta\;4861$ & \phs0.000 &
	  $100.0 \pm 4.5$ & $100.0 \pm 4.1$ &
	  $100.0 \pm 4.5$ & $100.0 \pm 6.2$ &
	  $100.0 \pm 4.2$ & $100.0 \pm 3.8$ 
\\
$[{\rm O\;III}]\;4959$ & $-0.026$ &
	  $189 \pm 14$ & $171 \pm 13$ &
	  $193 \pm 15$ & $169 \pm 25$ &
	  $158 \pm 14$ & $145 \pm 13$ 
\\
$[{\rm O\;III}]\;5007$ & $-0.038$ &
	  $546 \pm 19$ & $494 \pm 57$ &
	  $558 \pm 19$ & $487 \pm 54$ &
	  $466 \pm 19$ & $427 \pm 18$ 
\\
${\rm He\;I}\;5876$ & $-0.204$ &
	  $8.24 \pm 0.98$ & $7.45 \pm 0.89$ &
	  $7.4 \pm 1.3$ & $6.5 \pm 1.6$ &
	  $10.8 \pm 1.3$ & $10.0 \pm 1.2$
\\
${\rm H}\alpha\;6563$ & $-0.299$ &
	  $309.3 \pm 9.4$ & $286.0 \pm 9.4$ &
	  $316 \pm 10$ & $284 \pm 31$ &
	  $300 \pm 10$ & $284 \pm 13$
\\
$[{\rm N\;II}]\;6583$ & $-0.302$ &
	  $7.6 \pm 1.5$ & $6.9 \pm 1.4$ &
	  $10.8 \pm 3.4$ & $9.5 \pm 3.5$ &
	  $16.0 \pm 3.6$ & $14.8 \pm 3.4$
\\
$[{\rm S\;II}]\;6716,6731$ & $-0.320$ &
	  $39.6 \pm 3.4$ & $35.8 \pm 2.2$ &
	  $44.2 \pm 3.1$ & $37.8 \pm 7.8$ &
	  $49.9 \pm 5.9$ & $46.7 \pm 5.7$ 
\\
$[{\rm Ar\;III}]\;7136$ & $-0.375$ &
	  $10.6 \pm 1.5$ & $9.6 \pm 1.9$ &
	  $10.2 \pm 1.6$ & $8.9 \pm 2.0$ &
	  $11.9 \pm 1.7$ & $11.0 \pm 2.2$
\\[1mm]
\multicolumn{2}{c}{$F(\hbeta)$ (ergs s$^{-1}$ cm$^{-2}$)} & 
	  \multicolumn{2}{c}{$(8.84 \pm 0.40) \times 10^{-15}$} &
	  \multicolumn{2}{c}{$(2.85 \pm 0.13) \times 10^{-15}$} &
	  \multicolumn{2}{c}{$(4.27 \pm 0.18) \times 10^{-15}$} 
\\
\multicolumn{2}{c}{EW$_{\rm e}$(\hbeta) (\AA)} &
	  \multicolumn{2}{c}{$21.8 \pm 1.0$} &
	  \multicolumn{2}{c}{$13.9 \pm 0.64$} &
	  \multicolumn{2}{c}{$24.5 \pm 1.1$} 
\\
\multicolumn{2}{c}{Derived $E(B-V)$ (mag)} &
	  \multicolumn{2}{c}{$+0.008 \pm 0.109$} &
	  \multicolumn{2}{c}{$-0.007 \pm 0.112$} &
	  \multicolumn{2}{c}{$-0.013 \pm 0.108$} 
\\
\multicolumn{2}{c}{$c(\hbeta)$} &
	  \multicolumn{2}{c}{$+0.010$} &
	  \multicolumn{2}{c}{\nodata} &
	  \multicolumn{2}{c}{$-0.021$} 
\\
\multicolumn{2}{c}{Adopted $A_V$ (mag)} &
	  \multicolumn{2}{c}{$+0.02$} &
	  \multicolumn{2}{c}{0} &
	  \multicolumn{2}{c}{0}
\\
\multicolumn{2}{c}{EW$_{\rm abs}$ (\AA)} &
	  \multicolumn{2}{c}{$2.6$} &
	  \multicolumn{2}{c}{2} &
	  \multicolumn{2}{c}{$2.3$} 
\\[1mm]
\tableline
& & \multicolumn{2}{c}{B7} & 
\multicolumn{2}{c}{B8} & 
\\
\multicolumn{1}{c}{Wavelength (\AA)} &
\multicolumn{1}{c}{$f(\lambda)$} &
\multicolumn{1}{c}{$F$} & \multicolumn{1}{c}{$I$} &
\multicolumn{1}{c}{$F$} & \multicolumn{1}{c}{$I$} \\
\tableline
$[\rm{O\;II}]\;3727$ & $+0.325$ &
	  $271.4 \pm 7.3$ & $264 \pm 25$ &
	  $490 \pm 20$ & $513 \pm 86$ &
\\
$[{\rm Ne\;III}]\;3869$ & $+0.294$ & 
	  $60.9 \pm 6.5$ & $59.3 \pm 9.8$ &
	  \nodata & \nodata &
\\
${\rm H}\epsilon + {\rm He\;I}\;3970$ & $+0.269$ &
	  $27.8 \pm 4.8$ & $30.6 \pm 7.7$ &
	  \nodata & \nodata &
\\
${\rm H}\delta\;4101$ & $+0.232$ &
	  $28.3 \pm 3.0$ & $31.1 \pm 5.5$ &
	  $35.0 \pm 7.1$ & $37 \pm 12$ &
\\
${\rm H}\gamma\;4340$ & $+0.158$ &
	  $50.4 \pm 2.6$ & $52.1 \pm 6.1$ &
	  $34.1 \pm 5.4$ & $36 \pm 10$ &
\\
${\rm H}\beta\;4861$ & \phs0.000 &
	  $100.0 \pm 3.8$ & $100.0 \pm 5.3$ &
	  $100 \pm 10$ & $100 \pm 12$ &
\\
$[{\rm O\;III}]\;4959$ & $-0.026$ &
	  $116.5 \pm 8.7$ & $113 \pm 15$ &
	  $60.1 \pm 7.4$ & $59 \pm 14$ &
\\
$[{\rm O\;III}]\;5007$ & $-0.038$ &
	  $349 \pm 11$ & $340 \pm 34$ &
	  $207.6 \pm 9.7$ & $204 \pm 35$ &
\\
${\rm He\;I}\;5876$ & $-0.204$ &
	  $7.8 \pm 1.7$ & $7.6 \pm 2.1$ &
	  \nodata & \nodata &
\\
${\rm H}\alpha\;6563$ & $-0.299$ &
	  $288.5 \pm 9.2$ & $283 \pm 28$ &
	  $302 \pm 13$ & $286 \pm 49$ &
\\
$[{\rm S\;II}]\;6716,6731$ & $-0.320$ &
	  $33.6 \pm 3.6$ & $32.7 \pm 5.4$ &
	  \nodata & \nodata &
\\[1mm]
\multicolumn{2}{c}{$F(\hbeta)$ (ergs s$^{-1}$ cm$^{-2}$)} & 
	  \multicolumn{2}{c}{$(3.31 \pm 0.13) \times 10^{-15}$} &
	  \multicolumn{2}{c}{$(6.83 \pm 0.71) \times 10^{-16}$} 
\\
\multicolumn{2}{c}{EW$_{\rm e}$(\hbeta) (\AA)} &
	  \multicolumn{2}{c}{$71.8 \pm 4.2$} &
	  \multicolumn{2}{c}{$176 \pm 55$} 
\\
\multicolumn{2}{c}{Derived $E(B-V)$ (mag)} &
	  \multicolumn{2}{c}{$-0.012 \pm 0.102$} &
	  \multicolumn{2}{c}{$+0.041 \pm 0.172$} 
\\
\multicolumn{2}{c}{$c(\hbeta)$} &
	  \multicolumn{2}{c}{\nodata} &
	  \multicolumn{2}{c}{\nodata} 
\\
\multicolumn{2}{c}{Adopted $A_V$ (mag)} &
	  \multicolumn{2}{c}{0} &
	  \multicolumn{2}{c}{$+0.13$} 
\\
\multicolumn{2}{c}{EW$_{\rm abs}$ (\AA)} &
	  \multicolumn{2}{c}{2} &
	  \multicolumn{2}{c}{2}
\\
\tableline
\end{tabular}
\tablecomments{
See Table~\ref{table_data1} for comments.
}
\end{center}
\end{table}


\begin{table}
\scriptsize 
\tablenum{5c}
\begin{center}
\renewcommand{\arraystretch}{1.05} 
\caption{
Line ratios and properties for \hii\ regions C1, C2, C3, C4, and C6.
\vspace*{3mm}
\label{table_data3}
}
\begin{tabular}{rccccccc}
\tableline \tableline
& & \multicolumn{2}{c}{C1} & 
\multicolumn{2}{c}{C2} & 
\multicolumn{2}{c}{C3} \\
\multicolumn{1}{c}{Wavelength (\AA)} &
\multicolumn{1}{c}{$f(\lambda)$} &
\multicolumn{1}{c}{$F$} & \multicolumn{1}{c}{$I$} &
\multicolumn{1}{c}{$F$} & \multicolumn{1}{c}{$I$} &
\multicolumn{1}{c}{$F$} & \multicolumn{1}{c}{$I$} \\
\tableline
$[\rm{O\;II}]\;3727$ & $+0.325$ &
	  $332.5 \pm 7.1$ & $401 \pm 38$ &
	  $367.0 \pm 7.9$ & $402 \pm 38$ &
	  $425.8 \pm 8.6$ & $408 \pm 39$
\\
$[{\rm Ne\;III}]\;3869$ & $+0.294$ &
	  $27.4 \pm 1.4$ & $31.8 \pm 3.7$ &
	  $41.3 \pm 1.7$ & $44.2 \pm 4.8$ &
	  $67.7 \pm 3.2$ & $64.7 \pm 7.4$ 
\\
${\rm H}8 + {\rm He\;I}\;3889$ & $+0.289$ &
	  $11.0 \pm 1.2$ & $18.5 \pm 3.9$ &
	  $9.8 \pm 1.4$ & $18.0 \pm 4.7$ &
	  $8.6 \pm 2.5$ & $17.7 \pm 9.5$
\\
${\rm H}\epsilon + {\rm He\;I}\;3970$ & $+0.269$ &
	  $16.6 \pm 1.3$ & $24.3 \pm 3.9$ &
	  $13.0 \pm 1.4$ & $21.1 \pm 4.4$ &
	  $12.0 \pm 2.4$ & $21.0 \pm 7.6$ 
\\
${\rm H}\delta\;4101$ & $+0.232$ &
	  $17.6 \pm 1.4$ & $24.5 \pm 3.8$ &
	  $15.9 \pm 1.3$ & $23.3 \pm 3.9$ &
	  $15.6 \pm 2.1$ & $23.8 \pm 5.9$ 
\\
${\rm H}\gamma\;4340$ & $+0.158$ &
	  $43.0 \pm 1.3$ & $49.3 \pm 5.1$ &
	  $42.2 \pm 1.5$ & $48.0 \pm 5.2$ &
	  $42.6 \pm 2.6$ & $47.6 \pm 6.5$
\\
$[{\rm O\;III}]\;4363$ & $+0.151$ &
	  $< 3.50$ & $< 3.67$ &
	  $< 5.64$ & $< 5.67$ &
	  $< 10.4$ & $< 9.89$
\\
${\rm He\;I}\;4471$ & $+0.116$ &
	  $3.86 \pm 0.71$ & $3.97 \pm 0.96$ &
	  $2.97 \pm 0.65$ & $2.94 \pm 0.82$ &
	  \nodata & \nodata 
\\
${\rm H}\beta\;4861$ & \phs0.000 &
	  $100.0 \pm 4.0$ & $100.0 \pm 5.4$ &
	  $100.0 \pm 3.9$ & $100.0 \pm 5.4$ &
	  $100.0 \pm 4.1$ & $100.0 \pm 5.6$ 
\\
$[{\rm O\;III}]\;4959$ & $-0.026$ &
	  $113.8 \pm 7.3$ & $109 \pm 14$ &
	  $132.1 \pm 8.7$ & $126 \pm 16$ &
	  $167 \pm 12$ & $157 \pm 21$
\\
$[{\rm O\;III}]\;5007$ & $-0.038$ &
	  $334.7 \pm 9.5$ & $319 \pm 31$ &
	  $379 \pm 11$ & $359 \pm 36$ &
	  $467 \pm 15$ & $439 \pm 45$
\\
${\rm He\;I}\;5876$ & $-0.204$ &
	  $8.65 \pm 0.81$ & $7.5 \pm 1.2$ &
	  $6.52 \pm 0.87$ & $5.8 \pm 1.1$ &
	  $7.4 \pm 1.3$ & $6.9 \pm 1.6$
\\
${\rm H}\alpha\;6563$ & $-0.299$ &
	  $342.9 \pm 7.5$ & $286 \pm 27$ &
	  $326.1 \pm 8.2$ & $286 \pm 28$ &
	  $304.5 \pm 9.8$ & $286 \pm 30$
\\
$[{\rm N\;II}]\;6583$ & $-0.302$ &
	  $11.7 \pm 4.0$ & $9.7 \pm 3.9$ &
	  $17.7 \pm 5.4$ & $15.4 \pm 5.6$ &
	  $13.2 \pm 3.2$ & $12.2 \pm 3.6$ 
\\
${\rm He\;I}\;6678$ & $-0.314$ &
	  \nodata & \nodata &
	  \nodata & \nodata &
	  $5.8 \pm 2.0$ & $5.4 \pm 2.2$
\\
$[{\rm S\;II}]\;6716,6731$ & $-0.320$ &
	  $44.9 \pm 2.6$ & $43.2 \pm 5.3$ &
	  $49.9 \pm 3.0$ & $43.2 \pm 5.3$ &
	  $49.7 \pm 2.7$ & $46.0 \pm 5.5$ 
\\[1mm]
\multicolumn{2}{c}{$F(\hbeta)$ (ergs s$^{-1}$ cm$^{-2}$)} & 
	  \multicolumn{2}{c}{$(4.90 \pm 0.20) \times 10^{-15}$} &
	  \multicolumn{2}{c}{$(8.48 \pm 0.33) \times 10^{-15}$} &
	  \multicolumn{2}{c}{$(2.55 \pm 0.10) \times 10^{-15}$} 
\\
\multicolumn{2}{c}{EW$_{\rm e}$(\hbeta) (\AA)} &
	  \multicolumn{2}{c}{$64.5 \pm 3.8$} &
	  \multicolumn{2}{c}{$45.3 \pm 2.2$} &
	  \multicolumn{2}{c}{$31.6 \pm 1.5$} 
\\
\multicolumn{2}{c}{Derived $E(B-V)$ (mag)} &
	  \multicolumn{2}{c}{$+0.159 \pm 0.097$} &
	  \multicolumn{2}{c}{$+0.098 \pm 0.098$} &
	  \multicolumn{2}{c}{$+0.014 \pm 0.105$} 
\\
\multicolumn{2}{c}{$c(\hbeta)$} &
	  \multicolumn{2}{c}{\nodata} &
	  \multicolumn{2}{c}{\nodata} &
	  \multicolumn{2}{c}{\nodata} 
\\
\multicolumn{2}{c}{Adopted $A_V$ (mag)} &
	  \multicolumn{2}{c}{$+0.49$} &
	  \multicolumn{2}{c}{$+0.30$} &
	  \multicolumn{2}{c}{$+0.04$} 
\\
\multicolumn{2}{c}{EW$_{\rm abs}$ (\AA)} &
	  \multicolumn{2}{c}{2} &
	  \multicolumn{2}{c}{2} &
	  \multicolumn{2}{c}{2} 
\\[1mm]
\tableline
& & \multicolumn{2}{c}{C4} & 
\multicolumn{2}{c}{C6} \\ 
\multicolumn{1}{c}{Wavelength (\AA)} &
\multicolumn{1}{c}{$f(\lambda)$} &
\multicolumn{1}{c}{$F$} & \multicolumn{1}{c}{$I$} &
\multicolumn{1}{c}{$F$} & \multicolumn{1}{c}{$I$} \\
\tableline
$[\rm{O\;II}]\;3727$ & $+0.325$ &
	  $354.6 \pm 9.1$ & $352 \pm 56$ &
	  $310.0 \pm 5.9$ & $301.8 \pm 7.3$ &
\\
$[{\rm Ne\;III}]\;3869$ & $+0.294$ &
	  $91.3 \pm 7.8$ & $89 \pm 18$ &
	  $49.1 \pm 1.5$ & $47.8 \pm 1.6$ &
\\
${\rm H}8 + {\rm He\;I}\;3889$ & $+0.289$ &
	  \nodata & \nodata &
	  $18.9 \pm 1.3$ & $23.4 \pm 1.3$ &
\\
${\rm H}\epsilon + {\rm He\;I}\;3970$ & $+0.269$ &
	  \nodata & \nodata &
	  $25.3 \pm 1.3$ & $29.4 \pm 1.3$ &
\\
${\rm H}\delta\;4101$ & $+0.232$ &
	  \nodata & \nodata &
	  $20.3 \pm 1.5$ & $24.1 \pm 1.5$ &
\\
${\rm H}\gamma\;4340$ & $+0.158$ &
	  $23.2 \pm 2.9$ & $38 \pm 12$ &
	  $47.4 \pm 1.5$ & $49.6 \pm 1.5$ &
\\
$[{\rm O\;III}]\;4363$ & $+0.151$ &
	  \nodata & \nodata &
	  $3.4 \pm 1.2$ & $3.3 \pm 1.2$ &
\\
${\rm H}\beta\;4861$ & \phs0.000 &
	  $100.0 \pm 9.6$ & $100 \pm 12$ &
	  $100.0 \pm 3.7$ & $100.0 \pm 3.6$ &
\\
$[{\rm O\;III}]\;4959$ & $-0.026$ &
	  $188 \pm 10$ & $163 \pm 29$ &
	  $141 \pm 10$ & $137.3 \pm 9.7$ &
\\
$[{\rm O\;III}]\;5007$ & $-0.038$ &
	  $536 \pm 13$ & $464 \pm 73$ &
	  $404 \pm 14$ & $393 \pm 14$ &
\\
${\rm He\;I}\;5876$ & $-0.204$ &
	  \nodata & \nodata &
	  $7.87 \pm 0.80$ & $7.66 \pm 0.78$ &
\\
${\rm H}\alpha\;6563$ & $-0.299$ &
	  $350 \pm 10$ & $286 \pm 47$ &
	  $285.7 \pm 7.9$ & $279.7 \pm 8.6$ &
\\
$[{\rm N\;II}]\;6583$ & $-0.302$ &
	  $2.0 \pm 8.9$ & $1.6 \pm 7.4$ &
	  $6.4 \pm 6.7$ & $6.2 \pm 6.5$ &
\\
$[{\rm S\;II}]\;6716,6731$ & $-0.320$ &
	  $48.5 \pm 4.3$ & $38.5 \pm 8.1$ &
	  $37.5 \pm 1.9$ & $36.5 \pm 1.9$ &
\\
$[{\rm Ar\;III}]\;7136$ & $-0.375$ &
	  \nodata & \nodata &
	  $6.8 \pm 1.0$ & $6.6 \pm 1.3$ &
\\[1mm]
\multicolumn{2}{c}{$F(\hbeta)$ (ergs s$^{-1}$ cm$^{-2}$)} & 
	  \multicolumn{2}{c}{$(2.38 \pm 0.23) \times 10^{-15}$} &
	  \multicolumn{2}{c}{$(1.040 \pm 0.038) \times 10^{-14}$} &
\\
\multicolumn{2}{c}{EW$_{\rm e}$(\hbeta) (\AA)} &
	  \multicolumn{2}{c}{$14.0 \pm 1.4$} &
	  \multicolumn{2}{c}{$73.2 \pm 4.2$} &
\\
\multicolumn{2}{c}{Derived $E(B-V)$ (mag)} &
	  \multicolumn{2}{c}{$+0.092 \pm 0.164$} &
	  \multicolumn{2}{c}{$-0.023 \pm 0.098$} &
\\
\multicolumn{2}{c}{$c(\hbeta)$} &
	  \multicolumn{2}{c}{\nodata} &
	  \multicolumn{2}{c}{$-0.051 \pm 0.066$}
\\
\multicolumn{2}{c}{Adopted $A_V$ (mag)} &
	  \multicolumn{2}{c}{$+0.28$} &
	  \multicolumn{2}{c}{0} &
\\
\multicolumn{2}{c}{EW$_{\rm abs}$ (\AA)} &
	  \multicolumn{2}{c}{2} &
	  \multicolumn{2}{c}{$2.8 \pm 1.4$} 
\\
\tableline
\end{tabular}
\tablecomments{
See Table~\ref{table_data1} for comments.
}
\end{center}
\end{table}


\begin{deluxetable}{lccccccccc}
\tabletypesize{\small} 
\rotate
\tablenum{6a}
\renewcommand{\arraystretch}{1.1} 
\tablecolumns{10}
\tablewidth{0pt}
\tablecaption{
Ionic and total abundances.
\label{table_abund1}
}
\tablehead{
\colhead{Property} & 
\colhead{A1} & \colhead{A2} & \colhead{A3} & \colhead{A4} & 
\colhead{B1} & \colhead{B2} & \colhead{B3} & \colhead{B4} & \colhead{B5} 
}
\startdata
$T_e$(O$^{+2}$) (K) &
$< 20300$ & \nodata & $11400 \pm 1200$ & \nodata &
\nodata & \nodata & $12100 \pm 1200$ & $11400 \pm 1300$ & \nodata 
\\
$T_e$(O$^+$) (K) &
$< 17200$ & \nodata & $11000 \pm 1200$ & \nodata &
\nodata & \nodata & $11500 \pm 1200$ & $11000 \pm 1200$ & \nodata 
\\ 
O$^+$/H $(\times 10^5)$ & 
$> 2.1$ & \nodata & $6.5 \pm 2.7$ & \nodata &
\nodata & \nodata & $5.8 \pm 2.3$ & $7.9 \pm 3.5$ & \nodata 
\\
O$^{+2}$/H $(\times 10^5)$ &
$> 1.6$ & \nodata & $9.8 \pm 3.0$ & \nodata &
\nodata & \nodata & $9.3 \pm 2.7$ & $11.3 \pm 3.7$ & \nodata 
\\
O/H $(\times 10^5)$ & 
$> 3.7$ & \nodata & $16.2 \pm 4.0$ & \nodata &
\nodata & \nodata & $15.1 \pm 3.5$ & $19.2 \pm 5.1$ & \nodata 
\\
12$+$log(O/H) &
$> 7.57$ & \nodata & $8.21 \pm 0.10 \, (^{+0.11}_{-0.16})$ & \nodata &
\nodata & \nodata & $8.18 \pm 0.09 \, (^{+0.11}_{-0.14}) $ & 
$8.28 \pm 0.10 \, (^{+0.12}_{-0.16})$ & \nodata 
\\
%
%
12$+$log(O/H) M91$\,$\tablenotemark{a} &
8.21 & 8.33 & 8.15 & 8.11 & 8.25 & \nodata & 8.25 & 8.31 & 8.35
\\
12$+$log(O/H) P00$\,$\tablenotemark{b} &
8.14 & 8.11 & 7.96 & 7.93 & 8.18 & 8.29 & 8.02 & 8.08 & 8.13 
\\
\tableline
Ar$^{+2}$/H $(\times 10^7)$ &
\nodata & \nodata & $5.1 \pm 2.2$ & \nodata &
\nodata & \nodata & $5.8 \pm 2.9$ & $6.8 \pm 3.5$ & \nodata 
\\
ICF(Ar) &
\nodata & \nodata & 1.46 & \nodata &
\nodata & \nodata & 1.47 & 1.46 & \nodata 
\\
Ar/H $(\times 10^7)$ &
\nodata & \nodata & $7.5 \pm 3.2$ & \nodata &
\nodata & \nodata & $8.5 \pm 4.2$ & $9.8 \pm 5.1$ & \nodata 
\\
log(Ar/O) &
\nodata & \nodata & $-2.34 \pm 0.23$ & \nodata &
\nodata & \nodata & $-2.26 \pm 0.27$ & $-2.29 \pm 0.29$ & \nodata 
\\
N$^+$/O$^+$ $(\times 10^2)$ &
\nodata & \nodata & $2.2 \pm 1.6$ & \nodata &
\nodata & \nodata & $2.34 \pm 0.42$ & $1.34 \pm 0.28$ & \nodata 
\\
log(N/O) &
\nodata & \nodata & $-1.65 \pm 0.24$ & \nodata &
\nodata & \nodata & $-1.63 \pm 0.07$ & $-1.87 \pm 0.08$ & \nodata 
\\
Ne$^{+2}$/O$^{+2}$ &
\nodata & \nodata & $0.346 \pm 0.051$ & \nodata &
\nodata & \nodata & $0.356 \pm 0.058$ & $0.429 \pm 0.078$ & \nodata 
\\
log(Ne/O) &
\nodata & \nodata & $-0.460 \pm 0.063$ & \nodata &
\nodata & \nodata & $-0.448 \pm 0.070$ & $-0.367 \pm 0.078$ & \nodata 
\\
\enddata
\tablenotetext{a}{
\cite{mcgaugh91} bright-line calibration.
}
\tablenotetext{b}{
\cite{pilyugin00} bright-line calibration.
}
\tablecomments{
Direct oxygen abundances are shown with two uncertainties.
The first uncertainty is the formal uncertainty in the derivation.
In parentheses is the range of possible values, expressed by the
maximum and minimum values of the oxygen abundance.
}
\end{deluxetable}


\begin{deluxetable}{lcccccccc}
\tabletypesize{\small} 
\rotate
\tablenum{6b}
\renewcommand{\arraystretch}{1.1} 
\tablecolumns{9}
\tablewidth{0pt}
\tablecaption{
Ionic and total abundances (continued).
\label{table_abund2}
}
\tablehead{
\colhead{Property} & \colhead{B6} & \colhead{B7} & \colhead{B8} & 
\colhead{C1} & \colhead{C2} & \colhead{C3} & \colhead{C4} &
\colhead{C6} 
}
\startdata
$T_e$(O$^{+2}$) (K) &
$12900 \pm 1400$ & \nodata & \nodata & 
$< 12100$ & $< 13700$ & $< 16100$ & \nodata & $10900 \pm 1400$
\\
$T_e$(O$^+$) (K) &
$12000 \pm 1400$ & \nodata & \nodata & 
$< 11500$ & $< 12600$ & $< 14200$ & \nodata & $10600 \pm 1400$
\\
O$^+$/H $(\times 10^5)$ & 
$7.8 \pm 3.0$ & \nodata & \nodata & 
$> 8.4$ & $> 6.0$ & $> 4.0$ & \nodata & $8.9 \pm 4.7$
\\
O$^{+2}$/H $(\times 10^5)$ &
$6.8 \pm 1.9$ & \nodata & \nodata & 
$> 6.0$ & $> 4.8$ & $> 4.0$ & \nodata & $10.7 \pm 4.2$
\\
O/H $(\times 10^5)$ & 
$14.6 \pm 3.6$ & \nodata & \nodata & 
$> 14$ & $> 11$ & $> 8.0$ & \nodata & $19.6 \pm 6.3$
\\
12$+$log(O/H) &
$8.17 \pm 0.10 \, (^{+0.11}_{-0.14})$ & \nodata & \nodata & 
$> 8.16$ & $> 8.04$ & $> 7.90$ & \nodata & 
$8.29 \pm 0.12 \, (^{+0.14}_{-0.20})$
\\
%
%
12$+$log(O/H) M91$\,$\tablenotemark{a} &
\nodata & 8.08 & 8.31 & 8.26 & 8.30 & \nodata & 8.33 & 8.21
\\
12$+$log(O/H) P00$\,$\tablenotemark{b} &
8.23 & 7.94 & 8.46 & 8.18 & 8.18 & 8.20 & 8.12 & 8.03
\\
\tableline
Ar$^{+2}$/H $(\times 10^7)$ &
$6.2 \pm 3.0$ & \nodata & \nodata & \nodata &
\nodata & \nodata & \nodata & $5.1 \pm 2.7$
\\
ICF(Ar) &
1.49 & \nodata & \nodata & \nodata &
\nodata & \nodata & \nodata & 1.45
\\
Ar/H $(\times 10^7)$ &
$9.3 \pm 4.4$ & \nodata & \nodata & \nodata &
\nodata & \nodata & \nodata & $7.3 \pm 3.9$
\\
log(Ar/O) &
$-2.19 \pm 0.26$ & \nodata & \nodata & \nodata &
\nodata & \nodata & \nodata & $-2.41 \pm 0.32$
\\
N$^+$/O$^+$ $(\times 10^2)$ &
$2.37 \pm 0.57$ & \nodata & \nodata & \nodata &
\nodata & \nodata & \nodata & \nodata 
\\
log(N/O) &
$-1.63 \pm 0.09$ & \nodata & \nodata & \nodata &
\nodata & \nodata & \nodata & \nodata 
\\
Ne$^{+2}$/O$^{+2}$ &
$0.452 \pm 0.083$ & \nodata & \nodata & \nodata &
\nodata & \nodata & \nodata & $0.359 \pm 0.063$
\\
log(Ne/O) &
$-0.345 \pm 0.079$ & \nodata & \nodata & \nodata &
\nodata & \nodata & \nodata & $-0.444 \pm 0.075$
\\
\enddata
\tablenotetext{a}{
\cite{mcgaugh91} bright-line calibration.
}
\tablenotetext{b}{
\cite{pilyugin00} bright-line calibration.
}
\tablecomments{
See Table~\ref{table_abund1} for comments.
}
\end{deluxetable}


\begin{deluxetable}{ccc}
\tablenum{7}
\renewcommand{\arraystretch}{1.1}
\tablecolumns{3}
\tablewidth{0pt}
\tablecaption{
Neutral and nebular gas phase abundances.
\label{table_compare}
}
\tablehead{
\colhead{Property (dex)} & \colhead{\hi} & \colhead{\hii\ (mean)}
}
\startdata
12$+$log(O/H)  & \phs$7.43 \pm 0.3$ & \phs$8.21 \pm 0.05$ \\
12$+$log(N/H)  & \phs$5.73 \pm 0.3$ & \phs$6.46 \pm 0.08$ \\
log(N/O)       & $-1.70 \pm 0.4$ & $-1.75 \pm 0.06$ \\
12$+$log(Ar/H) & \phs$5.26 \pm 0.3$ & \phs$5.91 \pm 0.10$ \\
log(Ar/O)      & $-2.17 \pm 0.4$ & $-2.31 \pm 0.11$ \\
\enddata
\tablecomments{
Neutral gas (\hi) phase abundances are taken from \cite{heckman01};
they assign 0.3~dex uncertainties for the total abundances, and we
have derived approximate 0.4~dex uncertainties for the abundance
ratios. 
Nebular gas (\hii) phase abundances are taken from the present work.
}
\end{deluxetable}

\end{document}